\documentclass[aps,preprintnumbers,reprint,showkeys,eqsecnum,nofootinbib,floatfix,amsmath,asmfonts,latexsym,amssymb]{revtex4-1}
\usepackage{hyperref,graphics,slashed,color,multirow,float,amsfonts,bbold,mathtools}
\usepackage[T1]{fontenc} 
\usepackage{tikz}
\usetikzlibrary{positioning,arrows}
\usetikzlibrary{decorations.pathmorphing}
\usetikzlibrary{decorations.markings}
\usepackage{booktabs,tabularx,comment}

\graphicspath{{sections/figures/}}
\pdfoutput=1
\allowdisplaybreaks

\begin{document}
\preprint{IP/BBSR/2023-02}
\title{Radiative Neutrino Mass with Electroweak Scale Majorana Dark Matter in Scotogenic Model}
\author{Avnish}
\email[]{avnish@iopb.res.in} 
\author{Kirtiman Ghosh}
\email[]{kirti.gh@gmail.com}
\affiliation{Institute of Physics, Bhubaneswar, Sachivalaya Marg, Sainik School, Bhubaneswar 751005, India}%
\affiliation{Homi Bhabha National Institute, Training School Complex, Anushakti Nagar, Mumbai 400094, India}%
\date{\today}

\begin{abstract}

  \noindent Non-zero neutrino mass and dark matter cast a shadow over the success of the Standard Model (SM) of Particle Physics. The most straightforward extension of the SM to explain these two phenomena is the Scotogenic model, where the SM particle spectrum extends with three isospin singlet right-handed neutrinos and one doublet scalar while all of these being odd under $Z_2$ symmetry. In this work, we have considered the lightest right-handed neutrino as the dark matter candidate and freeze-out mechanism for producing observed dark matter relic density. The observed dark matter relic density, neutrino oscillation data and limits on the charged lepton flavor violation processes impose severe constraints on the model. After satisfying all the constraints, we study the collider signatures of the model at the proposed lepton collider experiments.
\end{abstract}
\maketitle

\tikzset{
  vector/.style={->,decorate, decoration={snake,amplitude=.4mm,segment length=2mm,post length=1mm}, draw},
  tes/.style={draw=black,postaction={decorate},
    decoration={snake,markings,mark=at position .55 with {\arrow[draw=black]{>}}}},
  provector/.style={decorate, decoration={snake,amplitude=2.5pt}, draw},
  antivector/.style={decorate, decoration={snake,amplitude=-2.5pt}, draw},
  fermion/.style={draw=black, postaction={decorate},decoration={markings,mark=at position .55 with {\arrow[draw=blue]{>}}}},
  fermionbar/.style={draw=black, postaction={decorate},
    decoration={markings,mark=at position .55 with {\arrow[draw=black]{<}}}},
  fermionnoarrow/.style={draw=black},
  scalar/.style={dashed,draw=black, postaction={decorate},decoration={markings,mark=at position .55 with {\arrow[draw=blue]{>}}}},
  scalarbar/.style={dashed,draw=black, postaction={decorate},decoration={markings,mark=at position .55 with {\arrow[draw=black]{<}}}},
  scalarnoarrow/.style={dashed,draw=black},
  electron/.style={draw=black, postaction={decorate},decoration={markings,mark=at position .55 with {\arrow[draw=black]{>}}}},
  bigvector/.style={decorate, decoration={snake,amplitude=4pt}, draw},
  particle/.style={thick,draw=blue, postaction={decorate},
    decoration={markings,mark=at position .5 with {\arrow[blue]{triangle 45}}}},
  gluon/.style={decorate, draw=black,
    decoration={coil,aspect=0.3,segment length=3pt,amplitude=3pt}}
}

\section{\label{sec:intro}Introduction}
The standard model (SM) of particle physics is the most successful phenomenological framework to describe and explain the non-gravitational fundamental interactions among elementary particles. However, there are several philosophically motivating issues and physically observed phenomenons which cannot be addressed in the framework of the SM. Non-zero tiny neutrino mass and their mixing established by neutrino oscillation observations and the absence of any particle candidate for dark matter (DM) to explain intriguing astronomical and cosmological observations are the most prominent ones to challenge the completeness of the SM. Although the current experimental data (from neutrino oscillation and scattering experiments) is inconclusive in determining the actual mechanism of neutrino mass generation, it is clear that massive neutrinos and dark matter are both part of nature; hence, they should be incorporated in the extensions of the SM. It would be fascinating if a single mechanism resolves these two most important outstanding puzzles.

The most intriguing aspect regarding neutrinos are the extreme minuteness of their masses and large mixing among the flavor eigenstates which are in stark contrast to other SM fermions ({\em i.e.,} charged leptons and quarks). While the simplest way to generate neutrino masses is to add right-handed neutrino fields to the SM particle content, it is hard to explain their extreme smallness. If neutrinos are Majorana particles, new physics beyond the electroweak scale could give rise to such a small neutrino mass via dimension-5 Weinberg operator \cite{Weinberg:1979sa}. At tree level, there are only three ways to generate the  Weinberg operator namely, type-I \cite{Minkowski:1977sc,Yanagida:1979as,Gell-Mann:1979vob,Mohapatra:1979ia}, type-II \cite{Magg:1980ut,Schechter:1980gr,Wetterich:1981bx,Lazarides:1980nt,Mohapatra:1980yp,Cheng:1980qt,FileviezPerez:2008jbu} and type-III \cite{Foot:1988aq}  seesaw mechanisms. However, in the tree level seesaws with ${\cal O}(1)$ couplings, tiny neutrino mass requires new physics at a very high scale ($\sim 10^{15}$ GeV) which is beyond the reach of collider experiments like the Large Hadron Collider (LHC), proposed electron-positron collider, {\em etc}. Whereas, models with radiatively generated neutrino mass \cite{Ma:1998dn,Babu:1988ki,FileviezPerez:2009ud,Babu:2002uu,Krauss:2002px,Cheung:2004xm,Ma:2006km,Ma:2007yx,Aoki:2008av,Aoki:2009vf,AristizabalSierra:2006gb,Cai:2017jrq} can be realized at the electroweak scale due to additional suppression from the loop integrals and hence, are easily testable at the collider experiments (For a review, see \cite{Cai:2017jrq}). Moreover, radiative neutrino mass models \cite{Ma:1998dn,Babu:1988ki,FileviezPerez:2009ud,Babu:2002uu,Krauss:2002px,Cheung:2004xm,Ma:2006km,Ma:2007yx,Aoki:2008av,Aoki:2009vf,AristizabalSierra:2006gb,Cai:2017jrq} usually require a ${\cal Z}_2$-symmetry to forbid the couplings, which lead to the generation of neutrino mass at tree level. The immediate consequence of this ${\cal Z}_2$ symmetry is the stability of the lightest ${\cal Z}_2$ odd particle, which if electrically neutral, can be a good candidate for the dark matter.

The SM symmetry group ($SU(3)_C\times SU(2)_L\times U(1)_Y$) supplemented by a ${\cal Z}_2$ symmetry and particle content extended with a new $SU(2)_L$ scalar doublet ($\Phi$), and two/three \footnote{Two/three generations of ${\cal Z}_2$-odd fermions are required to generate two/three non-vanishing neutrino masses. Note that one generation of them would not suffice to explain the experimental observation of at least two non-zero neutrino masses.} generations of heavy singlet fermions ($N_i$) both odd under the ${\cal Z}_2$ is the most straightforward (minimal) \cite{Ma:2006km} way to realize neutrino masses at one-loop level with a viable candidate for the DM in theory. The lepton number violating interactions, $-\frac{1}{2}\widetilde \lambda(H^\dagger \Phi)^2 -Y^{\alpha i}\overline{L_\alpha}\tilde{\Phi} N_i + {\rm h.c.}$, involving the ${\cal Z}_2$-odd exotics, SM lepton ($L_\alpha$) and Higgs ($H$) doublets result into Majorana neutrino masses, $m_\nu \sim {\cal O}\left( \frac{\widetilde\lambda}{32\pi^2} Y^TY \frac{v^2}{{\rm TeV}}\right)$\footnote{We have assumed TeV scale masses for the ${\cal Z}_2$-odd exotics. $32\pi^2$ suppression arises because of the loop.}, at the one-loop level where $v$ is the vacuum expectation value of the SM Higgs. Therefore, in this framework, $m_\nu \sim {\cal O}(0.1)$ eV can be realized for a wider range of the Yukawa couplings $Y^{\alpha i} \sim [{\cal O}(1)~-~{\cal O}(10^{-5})]$. Larger Yukawa couplings ($Y^{\alpha i}$), in general, result into enhanced rates for charged leptons flavor violating (CLFV) processes \cite{Toma:2013zsa}. Experimental limits \cite{MEG:2016leq,BaBar:2009hkt,SINDRUM:1987nra,Hayasaka:2010np,PhysRevLett.76.200,SINDRUMII:1998mwd,SINDRUMII:2006dvw,Kuno:2013mha,Pezzullo:2017iqq} on CLFV processes favour smaller $Y^{\alpha i}$s. However, in a scenario with ${\cal Z}_2$-odd exotic fermions being lighter than the ${\cal Z}_2$-odd scalars, smaller $Y^{\alpha i}$s result into smaller (co)annihilation cross-sections for the lightest ${\cal Z}_2$-odd particle and hence, give rise to a relic density (RD) which is larger than the WMAP/PLANK measurement \cite{WMAP:2012fli,Planck:2015fie}. To avoid this tension between the CLFV bounds \cite{MEG:2016leq,BaBar:2009hkt,SINDRUM:1987nra,Hayasaka:2010np,PhysRevLett.76.200,SINDRUMII:1998mwd,SINDRUMII:2006dvw,Kuno:2013mha,Pezzullo:2017iqq} and WMAP/PLANK \cite{WMAP:2012fli,Planck:2015fie} measured DM RD, several alternative (non-minimal) scenarios with additional exotic particles have been proposed in the literature \cite{ Borah:2017dqx, Borah:2021rbx, Borah:2021qmi, Sarma:2021acj,Soualah:2021xbn, Chao:2012sz, Sarazin:2021nwo, Abada:2021yot, Fiaschi:2018rky, DeRomeri:2021yjo, Avila:2019hhv, Beniwal:2020hjc}.

However, in this work, we stick to the minimal model \cite{Ma:2006km} with the exotic singlet fermions as the lightest ${\cal Z}_2$-odd particle and hence, the candidate for DM and search for a part of parameter space with relatively large Yukawa couplings but yet consistent with the CLFV bounds \cite{MEG:2016leq,BaBar:2009hkt,SINDRUM:1987nra,Hayasaka:2010np,PhysRevLett.76.200,SINDRUMII:1998mwd,SINDRUMII:2006dvw,Kuno:2013mha,Pezzullo:2017iqq}. The larger Yukawa couplings help explain the observed value of the DM RD and provide the opportunity to produce the ${\cal Z}_2$-odd singlet exotic fermions at the lepton colliders and study their signatures. After reproducing the experimentally measured low energy neutrino oscillation parameters ({\em i.e.,} neutrino mass square differences and mixing angles), WMAP/PLANK \cite{WMAP:2012fli,Planck:2015fie} measured DM RD and satisfying the CLFV constraints \cite{MEG:2016leq,BaBar:2009hkt,SINDRUM:1987nra,Hayasaka:2010np,PhysRevLett.76.200,SINDRUMII:1998mwd,SINDRUMII:2006dvw,Kuno:2013mha,Pezzullo:2017iqq}, we study the collider phenomenology of the ${\cal Z}_2$ odd exotics. The ${\cal Z}_2$-odd fermions, singlet under the SM gauge symmetry, cannot be produced at the Large Hadron Collider (LHC) experiment. However, due to the large Yukawa couplings involving the ${\cal Z}_2$-odd fermions and the SM leptons, they can be pair produced copiously at the lepton collider experiments. We investigate the signatures of exotic fermions at the proposed electron-positron collider experiments. On the other hand, the ${\cal Z}_2$-odd scalars, being doublet under the SM gauge group, can be pair produced at the LHC. We estimate the bounds on the masses of the ${\cal Z}_2$-odd scalars from the existing LHC studies \cite{ATLAS:2019lff} by the ATLAS/CMS collaborations with $\sqrt s=13$ TeV and run-II data.

\section{\label{sec:model}The Model}
To realize Weinberg operator \cite{Weinberg:1979sa} at the 1-loop level and obtain a cosmologically viable candidate for DM, the model incorporates three generations of new $SU(2)_L$ singlet fermions ($N_i$ with $i-1,2,3$) and a doublet scalar ($\Phi$) in the framework of the SM gauge symmetry supplemented by a $Z_2$-symmetry. The field content, along with their gauge quantum numbers, is summarized in the following: 
\begin{table}[h!]
  \centering
  \begin{tabular}{ |c||c|c|c|c|c|c||c|c| } 
    \hline\hline
    Symmetry & \multicolumn{6} {c||}{The SM fields} & \multicolumn{2} {c|}{Exotics}\\
    &$Q_L^\alpha$ & $u^\alpha_R$ & $d_R^\alpha$ & $L^\alpha_L$ & $e_R^{\alpha}$ & $H$ &$N_R^i$ &$\Phi$ \\
    \hline 
    $SU(3)_C$ &3 &3 &3 &1 & 1 & 1 &1 &1 \\ 
    $SU(2)_L$ &2  &1 &1 &2 & 1 & 2 &1 &1\\
    $U(1)_Y$ & $\frac{1}{6}$  &$\frac{2}{3}$  & $-\frac{1}{3}$  &$-\frac{1}{2}$  & $-1$ & $\frac{1}{2}$ &0 &$\frac{1}{2}$\\\hline
    $Z_2$ &+  &+ &+ &+ & + &+ &- &-\\
    \hline\hline
  \end{tabular}
  \caption{Field content of the model along with their gauge quantum numbers is presented where $\alpha,~ i=1,~2,~{\rm and}~3$ are the generation indices and the electric charges are determined by $Q=T_3+Y.$ Note that ${L_L^\alpha}^T$ = $\begin{pmatrix} \nu_L^\alpha & e_L^\alpha \end{pmatrix}$, ${Q_L^\alpha}^T$ = $\begin{pmatrix} u_L^\alpha & d_L^\alpha \end{pmatrix}$,    $H^T$ = $\begin{pmatrix} h^+ & \frac{v+h+i \eta}{\sqrt{2}} \end{pmatrix}$ and $\Phi^T$ = $\begin{pmatrix} \phi^+ & \frac{\phi_s+i \phi_p}{\sqrt{2}} \end{pmatrix}$.}
  \label{table:1}
\end{table}

\noindent The parts of the gauge invariant and renormalisable Lagrangian relevant for the generation of tiny non-vanishing neutrino masses at the one-loop level and DM phenomenology are given by 
\begin{eqnarray}
  \hfill
  &L& \ni - {\mu_\phi}^2 \Phi^\dagger \Phi -\lambda_0(\Phi^\dagger \Phi)^2-\lambda_1 (H^\dagger H)(\Phi^\dagger \Phi)-\lambda_2\left|H^\dagger \Phi\right|^2\nonumber\\&-&\frac{{\widetilde\lambda}}{2}\left[(H^\dagger \Phi)^2 +h.c. \right]+\frac{1}{2}M_{N_i}\bar N_i^C N_i-[Y^{\alpha i}\overline{L_L^\alpha}\tilde{\Phi} N_R^i + h.c.]\nonumber
  \label{eq:e5}
\end{eqnarray}
The scalar sector of inert two Higgs doublet model has been extensively studied in the literature\footnote{For brevity, we do not discuss different theoretical constraints (resulting from vacuum stability, perturbativity, unitarity, {\em etc.}) on the parameters of the scalar potential (see Ref.~\cite{Ahriche:2017iar}).} \cite{LopezHonorez:2006gr,Barbieri:2006dq,LopezHonorez:2010eeh,LopezHonorez:2010tb,Gustafsson:2012aj,Goudelis:2013uca,Garcia-Cely:2013zga,Arhrib:2013ela,Belyaev:2016lok,Tsai:2019eqi,Kalinowski:2020rmb}. To achieve the electroweak (EW) symmetry breaking and simultaneously preserve the $Z_2$ symmetry, the following assignments of VEVs are considered: $\langle H \rangle=\frac{v}{\sqrt{2}}$, where $v= 246$ GeV and  $ \langle \Phi \rangle=0$ \footnote{Note that $\Phi$ being odd under the $Z_2$, non-zero VEV for $\Phi$ will break the $Z_2$. However, to obtain a viable candidate for DM in the model, we need to preserve the $Z_2$ and hence, zero VEV is assigned to $\Phi$.}. In addition to the SM Higgs boson, the scalar sector of this model includes one CP - even ($\phi_s,~m^2_{\phi_s}=\mu^2_\phi+\frac{1}{2}\left( \lambda_1+\lambda_2 +{\widetilde\lambda} \right)v^2$), one CP - odd ($\phi_p,~m^2_{\phi_p}=\mu^2_\phi+\frac{1}{2}\left( \lambda_1+\lambda_2 -{\widetilde\lambda} \right)v^2$), and two singly charged scalars ($\phi^\pm,~m^2_{\phi^\pm}=\mu^2_\phi +\frac{1}{2}\lambda_1 v^2$), all odd under the $Z_2$. The fermionic sector includes three additional $Z_2$-odd fermions ($N_i$) singlet under the SM gauge symmetry with masses $M_{N_i}$. Note that the simultaneous presence of the Majorana mass terms and Yukawa terms involving $N_i$ and the $\widetilde \lambda$ term in the scalar potential violates the lepton number in the model and generates neutrino mass at the one-loop level, which we will discuss in the following.  

\begin{figure}
  \begin{center}
    \begin{tikzpicture}[line width=1.2 pt, scale=3.65,every node/.style={scale=1.0}]
      \draw[fermion,black,thin] (-1.0,0.0) --(-.50,0.0);
      \draw[fermion,black,thin] (-0.5,0) --(0,0);
      \draw[fermion,black,thin] (1.0,0) --(0.5,0);
      \draw[fermion,black,thin] (0.5,0) --(0.0,0);

      \draw[thin, dashed,->] (-.50,0) arc(180:90: 5 mm);
      \draw[thin, dashed, ->] (.50,0) arc(0:90: 5 mm);
      
      \draw[scalar,black,thin] (0.0,0.5)  --(0.7,1.0);
      \draw[scalar,black,thin] (0.0,0.5)  --(-0.7,1.0);    
      
      \node at (-0.8,-0.1) {$ \nu$};
      \node at (0.8,-0.1) {$ \nu$};
      \node at (-0.2,-0.1) {$ N_i$};
      \node at (0.2,-0.1) {$ N_i$};
      \node at (0.5,0.4) {$\phi_s, \phi_p$};
      \node at (-0.5,0.4) {$\phi_s, \phi_p$};
      \node at (-0.6,0.8) {$<H>$};
      \node at (0.6,0.8) {$<H>$};

      \node at (0.025,0.385)  {${\widetilde\lambda}$};  
    \end{tikzpicture}
  \end{center}
  \caption{\label{nu_mass}The neutrino mass generation at one loop level.}
\end{figure}
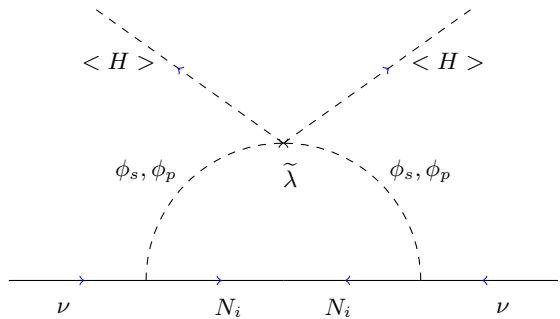

\subsection{\label{sec:numass}Neutrino Masses at One-Loop}
Weinberg operator \cite{Weinberg:1979sa} is generated at the one-loop level via the Feynman diagram depicted in Fig.~\ref{nu_mass}. Neutrinos get Majorana masses after the EWSB. The neutrino mass matrix resulting from the loop in Fig.~\ref{nu_mass} is given by\footnote{The approximate expression for the neutrino mass matrix holds for small mass splitting between the inert neutral scalars $\phi_s$ and $\phi_p$ {\em i.e.,}   $m_{\phi_s}^2-m_{\phi_p}^2 \sim \widetilde{\lambda} v^2 \ll M_{N_i}^2,~m_{\phi_s}^2~{\rm and}~m_{\phi_p}^2$.}
\begin{equation}
  M_\nu \simeq \frac{{\widetilde\lambda} v^2}{32\pi^2}\;\; \left(Y {\cal M}^{-1} Y^T\right),
  \label{mass_formula}
\end{equation}
where $[{\cal M}^{-1}]^{ij}= \delta^{ij} \frac{M_{N_i}^{-1}}{1-\chi_i}\left[1+\frac{{\rm ln}\chi_i}{1-\chi_i}\right]$, $\chi_i=\frac{\bar m^2}{M_{N_i}^2}$ and $\bar m^2=\frac{m_{\phi_s}^2+m_{\phi_p}^2}{2}$. Note that each of the following possibilities readily explains the smallness of the neutrino masses: (i) large masses for the exotic singlets, (ii) tiny Yukawa couplings, (iii) tiny quartic coupling ${\widetilde\lambda}$ in the scalar potential (iv) a combination of previous three. Scenarios with large masses for the $Z_2$-odd singlets are not testable at the collider experiments.  In a scenario with the exotic fermions being the lightest $Z_2$-odd particle, tiny Yukawa couplings result in large DM RD (see section \ref{relic} for details) already ruled out by WMAP/PLANK data \cite{WMAP:2012fli,Planck:2015fie}. In this work, we are interested in the third possibility, which allows TeV scale exotic scalars and fermions in the model with large Yukawa couplings ($\gtrsim {\mathcal O}(1)$) with the SM leptons. Large Yukawa couplings help explain the observed DM RD and give rise to interesting signatures at the lepton collider experiments. However, large Yukawa couplings also result in enhanced rates for the CLFV processes. The next section will address this tension between the large Yukawa couplings and bounds on the CLFV processes \cite{MEG:2016leq,BaBar:2009hkt,SINDRUM:1987nra,Hayasaka:2010np,PhysRevLett.76.200,SINDRUMII:1998mwd,SINDRUMII:2006dvw,Kuno:2013mha,Pezzullo:2017iqq}. Before that, we parameterize the Yukawa matrix $Y$ to explain the low-energy neutrino oscillation data in the next paragraph.

The neutrino mass formula in Eq.~\ref{mass_formula} contains more parameters (in particular, 15 independent parameters) at a higher scale than the nine parameters (3 neutrino masses, 3 mixing angles, and 3 phases) that determine the low-energy neutrino phenomenology. In our analysis, we have used Casas-Ibarra parametrization \cite{Casas:2006hf,Casas:2001sr} to find the general texture of the Yukawa matrix $Y$, which reproduces the measured mixing angles and mass spectrum of light neutrinos as follows  
\begin{eqnarray}
  Y= \frac{ 4 \sqrt{2} \pi}{\sqrt{{\widetilde\lambda}} v}\;\left(U_{\rm PMNS} \sqrt{\tilde{M_\nu}} R^T\sqrt{\cal{M}} \right)
  \label{eq:b7}
\end{eqnarray}
where, $\tilde{M_\nu}$ = diag($m_1,m_2,m_3$), with $m_1,m_2,m_3$ being light neutrino masses and $R$ is a complex orthogonal matrix, $R^T R=I$, of dimemsion $3\times 3$. Additionally, $U$ is Pontecorvo-Maki-Nakagawa-Sakata (PMNS) matrix~\cite{Maki:1962mu,Pontecorvo:1967fh} given by
{\tiny
\begin{eqnarray}
  U_{\rm PMNS}=\begin{pmatrix}
        c_{12}c_{13} &s_{12}c_{13} &s_{13}e^{-i\delta}\\
       -c_{23}s_{12} - s_{23}s_{13}c_{12}e^{i\delta} &c_{23}c_{12}-s_{23}s_{13}s_{12}e^{i\delta}  &s_{23}c_{13}\\
       s_{23}s_{12} - c_{23}s_{13}c_{12}e^{i\delta} &-s_{23}c_{12}-c_{23}s_{13}s_{12}e^{i\delta} &c_{23}c_{13}\\
  \end{pmatrix} P \nonumber
  \end{eqnarray}
}
where $c_{ij}={\rm cos} \theta_{ij},~s_{ij}={\rm sin} \theta_{ij}$, $\theta_{ij}$ with $i,j=1,2,3$ being mixing angles, $\delta$ is Dirac phase and the matrix $P={\rm diag}(e^{-i\alpha_1},e^{-i\alpha_2},1)$ contains the Majorana phases $\alpha_{1,2}$. The three light neutrino masses for normal mass hierarchy (NH) and inverted mass hierarchy (IH) scenarios are given by:
\begin{eqnarray}
  m_1~<~m_2=\sqrt{\Delta m^2_{21} + m^2_1}~<~m_3=\sqrt{\Delta m^2_{\rm 31} + m^2_1}~({\rm NH}) \nonumber
  \end{eqnarray}
\begin{eqnarray}
  m_3<m_1=\sqrt{|\Delta m^2_{32} +\Delta m^2_{\rm 21} -m^2_3|} < m_2=\sqrt{|\Delta m^2_{32}-m^2_3|}\nonumber \\
  \hspace{-5mm}({\rm IH}) \nonumber
\end{eqnarray}
where $\Delta m^2_{21}$, $\Delta m^2_{31}$, and $\Delta m^2_{32}$ are the neutrino mass squared differences measured from the neutrino oscillation experiments. The numerical values of the neutrino oscillation parameters, namely the mass square differences and the mixing angles used in our analysis, are presented in Table\ref{table:neut_osc}. For the sake of simplicity, the Dirac
and Majorana phases in the PMNS matrix are assumed to be zero in this work.
\begin{table}[ht!]
  \centering
  \scalebox{.98}{\begin{tabular}{|c|c|c|}\hline
      Parameters &NH &IH \\\hline
      $\Delta m^2_{21}/10^{-5} {\rm eV}^2$ &$6.80 \rightarrow 8.02$ &$6.80 \rightarrow 8.02$\\  
      $\Delta m^2_{3l}/10^{-3} {\rm eV}^2$ &$2.399 \rightarrow 2.544$ &$-2.6 \rightarrow -2.369$\\
      ${\rm sin}^2 \theta_{12}$ &$0.272 \rightarrow 0.346$ &$0.272 \rightarrow 0.346$ \\  
      ${\rm sin}^2 \theta_{23}$ &$0.418 \rightarrow 0.613$ &$0.435 \rightarrow 0.616$ \\
      ${\rm sin}^2 \theta_{13}$ &$0.01981 \rightarrow 0.02436$ &$0.02006 \rightarrow 0.02452$ \\\hline
   \end{tabular}}
  \caption{\label{table:neut_osc}The oscillation parameters in $3\sigma$ range for both NH and IH from the global analysis for the neutrino oscillation parameters with three light active neutrinos~\cite{Esteban:2016qun}. Point to note that $\Delta m^2_{3l}$ represents: $\Delta m^2_{31}$ for NH and  $\Delta m^2_{32}$ for IH.}
\end{table}

\subsection{\label{sec:cLFV}Constraints from Lepton Flavor Violations and the texture of Yukawa and $Z_2$--odd fermion masses}
In different neutrino mass models, flavour mixing in the neutrino sector, in general, leads to flavour violation in the charged lepton sector, which contributes to the observables like the charged lepton flavour violating decays  ($l_\alpha \rightarrow l_\beta \gamma$), $\mu$--$e$ conversion in nuclei e.t.c. These charged lepton flavour-violating observables are tightly constrained from different charged lepton-violation experiments. Therefore, in neutrino mass models, the parameters responsible for generating neutrino masses and mixings usually receive stringent constraints from the lepton flavour violation observables. In the present model, the Yukawa interactions involving the $Z_2$--odd singlet fermions ($N_R^i$), scalar ($\Phi$) and the SM lepton doublets ($L_L^\alpha$) lead to the charged lepton flavour violating decays as well as flavour conversion in the nucleus at one-loop level \cite{Toma:2013zsa}. The appropriate analytical expression for the branching ratio of the loop-induced flavour violating decay of the SM charged leptons ($l_\alpha \rightarrow  l_\beta$ $\gamma$) is given by \cite{Ma:2001mr,Kubo:2006yx},
\begin{eqnarray}
\hspace{-5mm}
 {\rm BR}(l_\alpha \rightarrow l_\beta \gamma)=\frac{3 \alpha_{em} v^4}{32 \pi m^4_{\phi^\pm}} \left|\left (Y \widetilde{{\cal{F}}}  Y^\dagger\right)^{\alpha\beta} \right|^2
\label{eq:a1}
\end{eqnarray}
where, $\widetilde{{\cal{F}}}$ is a $3\times3$ diagonal matrix defined as: $\widetilde{{\cal{F}}}={\rm diag}\left( {\cal{F}}(\xi_1), {\cal{F}}(\xi_2), {\cal{F}}(\xi_3)\right)$ with $\xi_i=\frac{M^2_{N_i}}{m^2_{\phi^\pm}}$ and ${\cal{F}}(z)=\frac{1-6z+3z^2+2z^3-6z^2 ln z}{6(1-z)^4}$, is a monotonically decreasing function which can be approximated as:
\begin{equation}
  {\cal{F}}_i(z)=\begin{cases}
    \frac{1}{3z}, & \text{for $z \gg 1$}.\\
    \frac{1}{6}, & \text{for $z \ll 1$}.
  \end{cases}
\end{equation}
Similar analytical expressions for other lepton flavour-violating processes, like the $\mu$--$e$ conversion rate in nuclei \cite{Vicente:2014wga,Ibarra:2016dlb,Toma:2013zsa}, branching ratios of other lepton flavour violating decays like $\mu^+ \to e^+ e^- e^+$, $\tau^+ \to e^+ e^- e^+$, $\tau^+ \to \mu^+ \mu^- \mu^+$ e.t.c., are available in the literature.
Lepton flavour-violating observables, being one of the most promising hints for the physics beyond the SM, are being extensively studied by different experimental collaborations in different LEV observables. Null results from all those experiments result into stringent bounds on the rate of different LFV processes. A list of upper bounds (present bound as well as future sensitivity for some of the observables) on different LEV observables (first column of Table~\ref{table:A2}) resulting from different experiments is presented in the second column of Table~\ref{table:A2}.

\begin{table*}[ht!]
  \centering
  \scalebox{1.0}{\begin{tabular}{|c|c|c|}\hline
      BP1 (NH) &BP2 (NH) &BP3 (IH) \\\hline
      \multicolumn{3}{|c|}{$\mu_\phi$ = 1 TeV, $\lambda_1$ = 0.004, $\lambda_2$ = 0.005}\\\hline
      $m_1$ = $4.16 \times 10^{-11}$, $M_{N_1} = 200$ GeV
      &$m_1$ = $3.02 \times 10^{-11}$, $M_{N_1} = 500$ GeV
      &$m_3$ = $6.6 \times 10^{-11}$, $M_{N_3} = 200$ GeV\\

      $\widetilde{\lambda}$ = $ 4.6\times 10^{-10}$
      &$\widetilde{\lambda}$ = $2.84 \times 10^{-10}$
      &$\widetilde{\lambda}$ = $2.47 \times 10^{-10}$\\\hline
      
      $M_{N_2}$ = 204.01 GeV, $M_{N_3}$ = 312.5 GeV
      & $M_{N_2}$ = 517.89 GeV, $M_{N_3}$ = 916 GeV
      & $M_{N_1}$ = 253.18 GeV, $M_{N_2}$ = 254.46 GeV\\

      $m_{\nu_2}$ =  $ 4.24 \times 10^{-11}$ GeV,
      & $m_{\nu_2}$ =  $3.13 \times 10^{-11}$ GeV,
      & $m_{\nu_1}$ =  $ 8.30\times 10^{-11}$ GeV, \\
      
      $m_{\nu_3}$ =  $6.43 \times 10^{-11}$ GeV, & $m_{\nu_3}$ =  $5.75 \times 10^{-11}$ GeV &$m_{\nu_2}$ =  $ 8.34 \times 10^{-11}$ GeV,\\

      $Y$=$\begin{pmatrix}
        1.37 &0.84 &0.24\\
        -0.77 &0.98 &1.07\\ 
        0.40 &-0.98 &1.27\\
      \end{pmatrix}$
      &
      $Y$ = $\begin{pmatrix}
        1.05 &0.65 &0.21\\
         -0.59 &0.76 &0.94\\
         0.31 &-0.76 &1.12\\
      \end{pmatrix}$
      &
       $Y$ = $\begin{pmatrix}
        2.39 &1.46 &0.4 \\
        -1.34 &1.67 &1.82\\
         0.71 &-1.75 &2.07\\
      \end{pmatrix}$\\

      $U$=$\begin{pmatrix}
        0.845 &0.516 &0.141\\
        -0.476 &0.603 &0.640\\ 
        0.246 &-0.608 &0.755\\
      \end{pmatrix}$

      &$U$=$\begin{pmatrix}
        0.845 &0.516 &0.141\\
        -0.476 &0.603 &0.640\\ 
        0.246 &-0.608 &0.755\\
      \end{pmatrix}$
      &$U$=$\begin{pmatrix}
        0.844 &0.516 &0.145\\
        -0.474 &0.591 &0.653\\ 
        0.251 &-0.620 &0.744\\
      \end{pmatrix}$\\\hline
  \end{tabular}}
  \caption{\label{table:A1}The benchmark points.}
\end{table*}

\begin{table*}[ht!]
  \centering
  \scalebox{1.155}{ \begin{tabular}{|c|c|c|c|c|}
      \hline
      Observables & Experimental limits & Estimate for BP1 & Estimate for BP2 & Estimate for BP3\\
      \hline
      BR($\mu^+ \rightarrow e^+ \gamma$) & $4.2 \times 10^{-13}$ \cite{MEG:2016leq} &$1.5 \times 10^{-20}$  &$9.86 \times 10^{-22}$ &$1.08 \times 10^{-20}$\\
      \hline
      BR($\tau^+ \rightarrow e^+ \gamma$) & $3.3 \times 10^{-8}$ \cite{BaBar:2009hkt} &$3.17 \times 10^{-21}$  &$3.76 \times 10^{-22}$ &$2.14 \times 10^{-20}$\\
      \hline
      BR($\tau^+ \rightarrow \mu^+ \gamma$) & $4.4 \times 10^{-8}$ \cite{BaBar:2009hkt} &$7.13 \times 10^{-21}$  &$6.58 \times 10^{-22}$ &$2.20 \times 10^{-19}$ \\
      \hline
      BR($\mu^+ \rightarrow e^+ e^- e^+ $) & $1.0 \times 10^{-12}$ \cite{SINDRUM:1987nra} &$6.19 \times 10^{-16}$  &$7.75 \times 10^{-13}$ &$7.88 \times 10^{-14}$\\
      \hline
      BR($\tau^+ \rightarrow e^+ e^- e^+ $) & $2.7 \times 10^{-8}$ \cite{Hayasaka:2010np}  &$3.23 \times 10^{-17}$ &$1.15 \times 10^{-13}$ &$2.96 \times 10^{-14}$\\
      \hline
      BR($\tau^+ \rightarrow  \mu^+ \mu^- \mu^+$) & $2.1 \times 10^{-8}$ \cite{Hayasaka:2010np} &$4.89 \times 10^{-14}$   &$4.49 \times 10^{-12}$  &$3.68 \times 10^{-14}$ \\
      \hline
      CR($\mu - e, Pb$) &  $4.6 \times 10^{-11}$ \cite{PhysRevLett.76.200} & $6.9 \times 10^{-15}$ &$3.85 \times 10^{-14}$ &$9.05 \times 10^{-15}$\\
      \hline
      CR($\mu - e, Ti$) & $1.7 \times 10^{-12}$ \cite{SINDRUMII:1998mwd} &$8.95 \times 10^{-15}$ &$4.99 \times 10^{-14}$ &$1.17 \times 10^{-14}$\\
      \hline
      CR($\mu - e, Au$) & $7.0 \times 10^{-13}$ \cite{SINDRUMII:2006dvw} &$7.34 \times 10^{-15}$ &$4.09 \times 10^{-14}$ &$9.63 \times 10^{-15}$\\
      \hline
      CR($\mu - e, Al$) & $1.0 \times 10^{-16}$ \cite{Kuno:2013mha, Pezzullo:2017iqq} &$4.97 \times 10^{-15}$ &$2.77 \times 10^{-14}$ &$6.52 \times 10^{-15}$ \\
      \hline
      $\Omega_{\rm DM} h^2$ & $0.119$ \cite{WMAP:2012fli} & 0.118 &0.119 &0.119\\
      \hline
  \end{tabular}}
  \caption{\label{table:A2}The model predictions for the LFV and dark matter observables corresponding to the three BPs listed in table~\ref{table:a1} are presented in columns 3, 4, and 5.  Corresponding experimental limits are given in column 2. $\Omega_{\rm DM} h^2$ is the dark matter relic density.}
\end{table*}

All new physics scenarios beyond the SM must be consistent with the upper bound on the LFV observables listed in Table~\ref{table:A2}. Note that the limit  BR($\mu \rightarrow e \gamma) \leq 4.2 \times 10^{-13}$ from MEG experiment leads to the most stringent constraints on the BSM scenarios. In the context of the present model, the three following scenarios result in suppressed branching ratios for $\mu \rightarrow e \gamma$ given in Eq.~\ref{eq:a1}. {\em Scenario I:}  BR($\mu \rightarrow e \gamma$) is suppressed by the masses of the $Z_2$--odd particles in the loop. Therefore, large masses ($\sim$ few hundreds of TeVs for the Yukawa coupling $\sim~{\cal{O}}(1)$) of the $Z_2$--odd particles automatically suppress the $\mu \rightarrow e \gamma$ rate. However, such scenarios are not testable in the collider experiment. {\em Scenario II:} Smaller values of the Yukawa couplings can also suppress the LFV branching ratios and conversion rates under the experimental upper bounds. Note that the same Yukaua couplings play a crucial role in determining the dark matter annihilation cross-section and hence, dark matter relic density in a scenario with a $Z_2$--odd singlet fermion as a candidate for the dark matter. Smaller Yukawa couplings result in smaller dark matter annihilation cross section and hence, predict relic densities that are inconsistent with the experimentally measured values from WMAP\cite{WMAP:2012fli} and Planck\cite{Planck:2015fie}. {\em Scenario III:}  For TeV scale masses of the $Z_2$--odd particles and ${\cal{O}}(1)$ Yukawa couplings, it may be possible to choose the texture of the Yukawa matrix ($Y$) and the masses of the $Z_2$--odd particles in such a way that the off-diagonal elements of $\left (Y \widetilde{{\cal{F}}}  Y^\dagger\right)$ vanishes. Such a scenario is phenomenologically interesting because, with the large Yukawa couplings, it easily explains the WMAP\cite{WMAP:2012fli} and Planck\cite{Planck:2015fie} measured relic density on one hand. On the other hand, the exotic TeV scale particles give rise to the possibility of probing the signatures of this model at the collider experiments. Therefore, in this work, we focused on {\em Scenario III}. In the following, we obtained the texture of Yukawa and $Z_2$--odd fermion masses that result in suppressed LFV rates for relatively large Yukawa and TeV scale masses for the $Z_2$--odd fermions. 

To suppress the  LFV branching ratio  of $\mu \rightarrow e \gamma$ in Eq.~\ref{eq:a1}, {\em Scenario III} requires suppressed or vanishing off-diagonal elements for $\left (Y \widetilde{{\cal{F}}}  Y^\dagger\right)$. Our aim here is to obtain the particular texture of the Yukawa matrix ($Y$) and the masses of the $Z_2$-odd right-handed neutrinos ($M_{N_1},~M_{N_2}$ and $M_{N_3}$) that result into
\begin{equation}
    Y \widetilde{{\cal{F}}}  Y^\dagger~=~D,
    \label{req1}
\end{equation}
where, $D~=~{\rm diag}(d_1,d_2,d_3)$ is an arbitrary diagonal matrix. Note that the neutrino oscillation data, namely the light neutrino masses (the $M_\nu$ matrix in Eq.~\ref{eq:b7})and mixing (the PMNS matrix $U$ in Eq.~\ref{eq:b7}), partially determines the texture of the Yakawa matrix $Y$ (see Eq.~\ref{eq:b7}). Substituting the general texture of the Yukawa matrix $Y$ (from Eq.~\ref{eq:b7}), which reproduces the measured light neutrino masses and mixings in Eq.~\ref{req1}, we obtain
\begin{equation}
U U^\dagger~=~I_{3\times 3}
\label{req2}
\end{equation}
where, 
\begin{equation}
    U~=~\frac{4\sqrt{2}\pi}{\sqrt{\tilde \lambda}v}\sqrt{D} U_{\rm PMNS} \sqrt{\tilde{M_\nu}} R^T \sqrt{{\cal{M}}} \sqrt{\tilde{\cal{F}}}.
    \label{req3}
\end{equation}
Given the fact that $U_{\rm PMNS}$ is a unitary matrix and already determined from the neutrino oscillation data, $R$ is an arbitrary complex orthogonal matrix, and the other matrices in Eq.~\ref{req3} are diagonal, Eq.~\ref{req2} will be satisfied under the condition that both  $\sqrt{D}$ and $\left(\sqrt{\tilde{M_\nu}} R^T \sqrt{{\cal{M}}} \sqrt{\tilde{\cal{F}}}\right)$ are proportional to some unitary matrices. Therefore, $D$, being a diagonal matrix, $D$ must be proportional to $I_{3\times 3}$. On the other hand, $\tilde{M_\nu},~{\cal{M}}~{\rm and}~\tilde{\cal{F}}$ being diagonal, the only possible solution for $R$ and the masses of the $Z_2$-odd right-handed neutrinos which leads to a unitary $\left(\sqrt{\tilde{M_\nu}} R^T \sqrt{{\cal{M}}} \sqrt{\tilde{\cal{F}}}\right)$ is presented in the following:
\begin{equation}
    R~\propto~I_{3\times 3}~~ {\rm and}~~ \left(\sqrt{\tilde{M_\nu}} \sqrt{{\cal{M}}} \sqrt{\tilde{\cal{F}}}\right)~\propto~I_{3\times3}
\end{equation}
Neutrino oscillation data, along with the requirement $R~\propto~I_{3\times 3}$, completely determine the Yakawa matrix $Y$. Whereas, $\left(\sqrt{\tilde{M_\nu}} \sqrt{{\cal{M}}} \sqrt{\tilde{\cal{F}}}\right)~\propto~I_{3\times3}$ results into the following relations among the $Z_2$-odd right-handed neutrino masses:
\begin{equation}
    {\cal{G}}\left(M_{N_1}\right) m_1 = {\cal{G}}\left(M_{N_2}\right) m_2 = {\cal{G}}\left(M_{N_3}\right) m_3,
    \label{req4}
\end{equation}
where,
\begin{equation}
  {\cal{G}}\left(m\right)~=~\frac{m}{\bar m^2 -m^2} \left[1 - \frac{m^2}{\bar m^2 -m^2} ln(\frac{\bar m^2}{m^2})\right] {\cal{F}}\left(\frac{m^2}{m^2_{\phi^\pm}}\right) \nonumber
\end{equation}


Neutrino oscillation data and the constraints for LFV observables make this scenario very predictive. While the texture of the Yakawa matrix $Y$ is completely determined for $R=I_{3\times3}$, $\tilde \lambda$ controls the strength of the Yukawa couplings. The texture of the $Z_2$-odd fermion masses is determined by Eq.~\ref{req4}, rendering the freedom to choose only one $Z_2$-odd fermion mass free parameter. Without knowledge about the absolute masses of the three light neutrinos, the lightest neutrino Mass remains an additional parameter in the motel. Therefore, the phenomenology of this model is essentially determined by three parameters, namely, $\tilde \lambda$, $M_{N_1}$, and $m_1(m_3)$ for NH(IH). With such a limited number of parameters, the model gives rise to interesting predictions which can be tasted at the collider experiments. Before going into collider phenomenology, we discuss the technical details of our numerical implementation and the resulting theoretical (from perturbativity of the Yukawa couplings) and experimental (from LFV observables and dark matter relic density) constraints in the following sections.

\begin{figure*}[]
  \centering 
  \includegraphics[width=0.48\textwidth,scale=1]{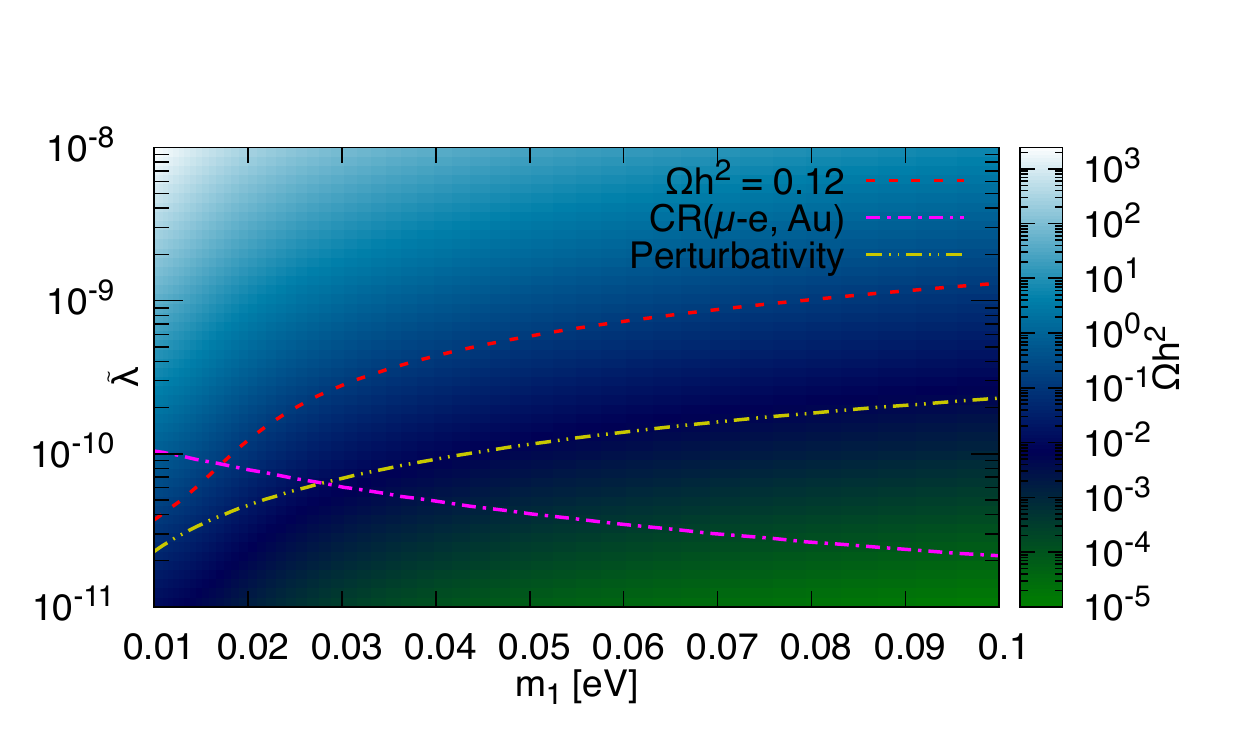}
  \includegraphics[width=0.48\textwidth,scale=1]{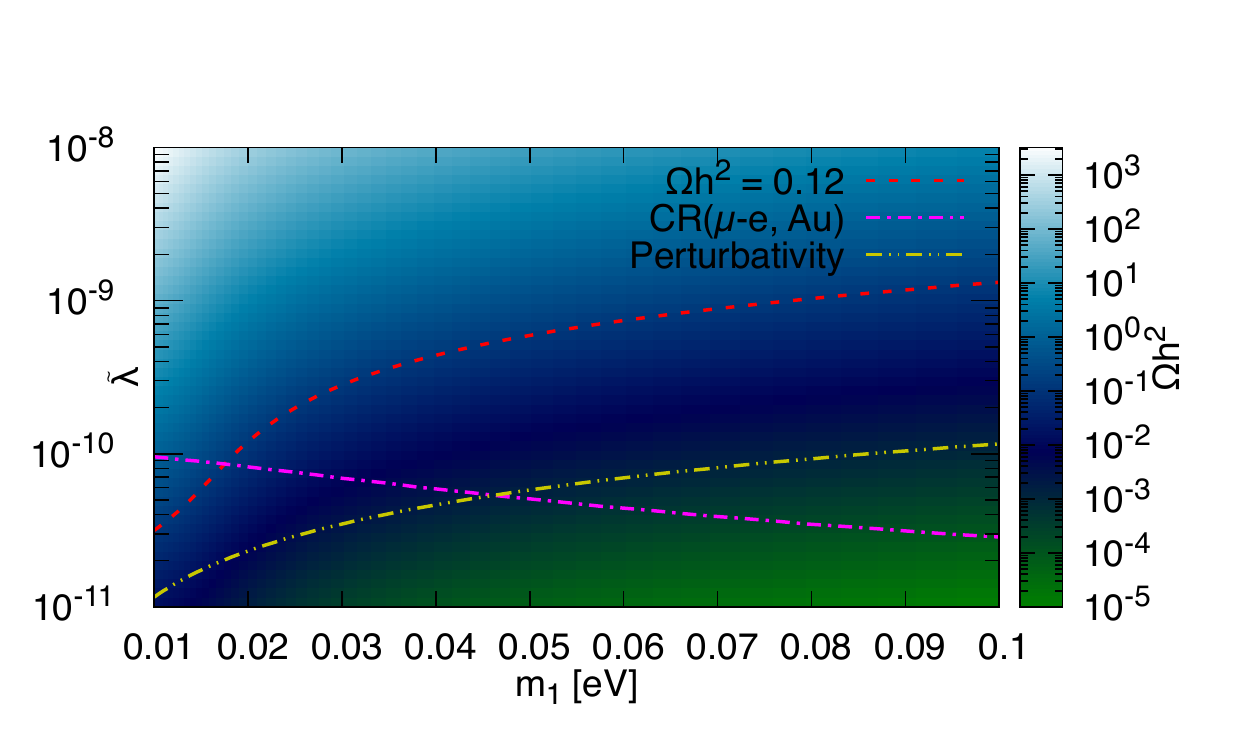}
  \includegraphics[width=0.48\textwidth,scale=1]{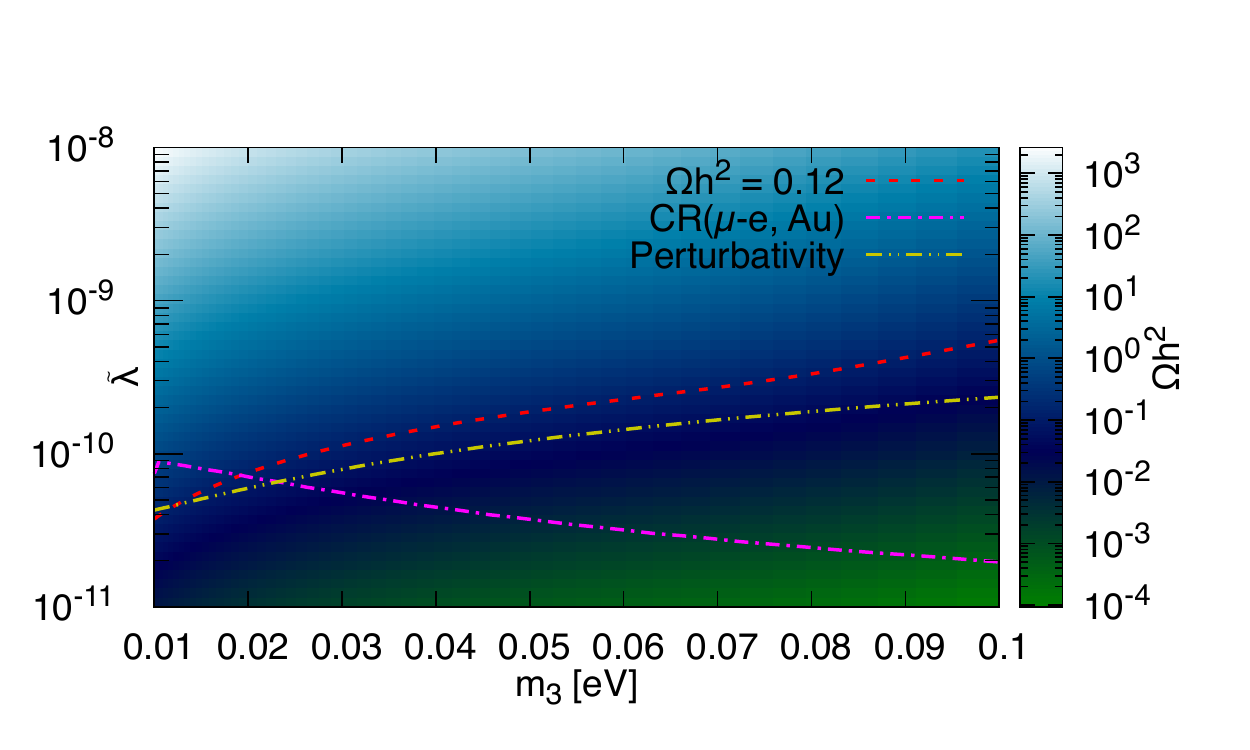}
  \includegraphics[width=0.48\textwidth,scale=1]{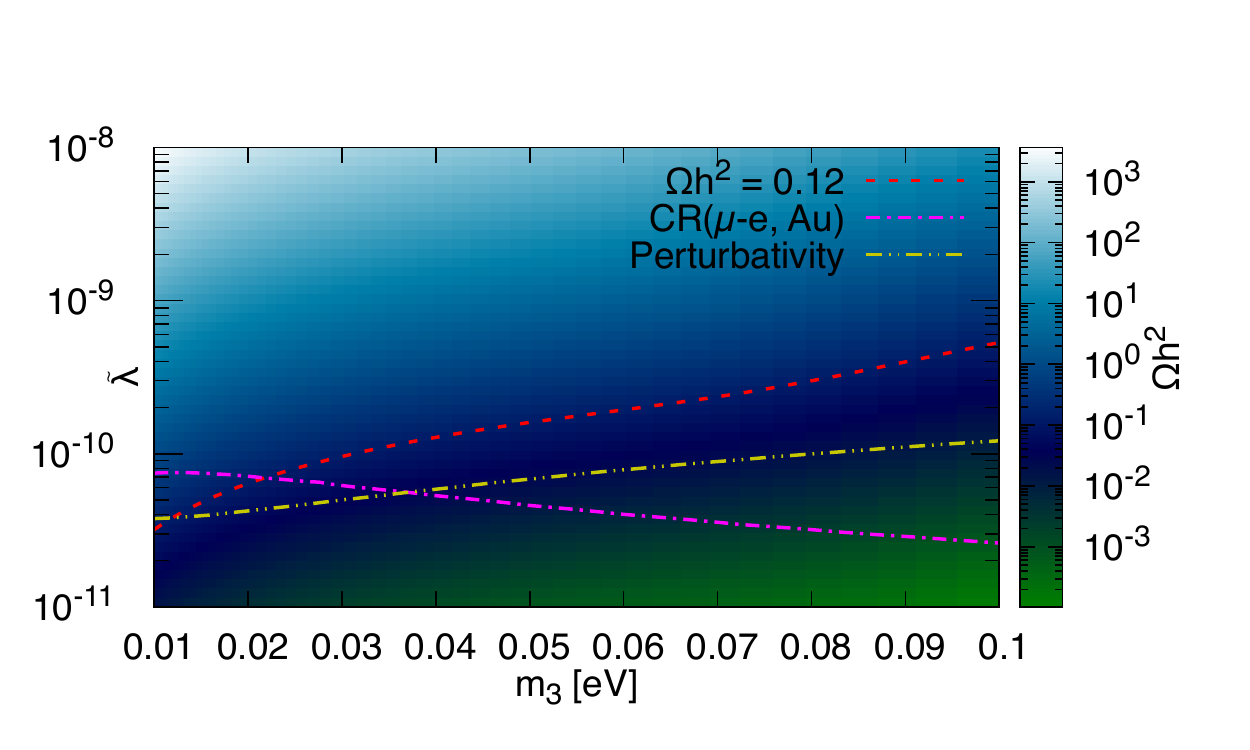}
  \caption{\label{fig:ex1}{The model prediction for the dark matter relic density (by the color gradient) is presented as a function of $m_1$ and $\widetilde{\lambda}$ for the normal mass hierarchy of the light neutrino masses and two fixed masses for the lightest $Z_2$-odd fermion (the dark matter candidate), namely  $M_{N_1}=$ 200 GeV (top left panel) and 500 GeV (top right panel). Similarly, the model prediction for the dark matter relic density is shown as a function of $m_3$ and $\widetilde{\lambda}$ for the inverted mass hierarchy of the light neutrino masses and two fixed masses for the lightest $Z_2$-odd fermion (the dark matter candidate), namely  $M_{N_3}=$ 200 GeV (bottom left panel) and 500 GeV (bottom right panel). For all plots, the points falling on the dashed line (red), dashed line with one dot (violet), and dashed line with two dots (yellow) correspond to the parameter values which yield WMAP\cite{WMAP:2012fli} and Planck\cite{Planck:2015fie} measured value for the dark matter relic density, CR$(\mu-e, Au) = 7.0 \times 10^{-13}$ and perturbativity limit of the Yukawa couplings ($Y_{11}=3$) respectively. We have considered $\mu_{\phi}=1$ TeV, $\lambda_{1} = 0.004$ and $\lambda_{2}= 0.005$ to generate these plots.}}
\end{figure*}

\section{\label{sec:NM}Numerical Implimentation}
After discussing the theoretical framework and the particular scenario, we are interested in; we are now equipped to study the phenomenology of the scenario in the context of LFV, dark matter, and collider experiments. To compute the numerical values of different observables, we have implemented the model in {\em SARAH-4.14.4}~\cite{Staub:2013tta,Staub:2015kfa} to produce {\em SPheno}~\cite{Porod:2003um,Porod:2011nf}, {\em micrOMEGAs-5.2.1}~\cite{Belanger:2018ccd} source modules and model {\em UFO}~\cite{Degrande:2011ua} file for {\em MadGraph5-2.6.7}~\cite{Alwall:2014hca}. {\em SPheno}~\cite{Porod:2003um,Porod:2011nf}, {\em micrOMEGAs-5.2.1}~\cite{Belanger:2018ccd} source modules are used for calculating the model predictions for the LFV observables and dark matter relic density. The productions of the $Z_2$-odd exotics at different collider environments are simulated in  {\em MadGraph5-2.6.7}~\cite{Alwall:2014hca} using the model {\em UFO}s.

\begin{figure*}[tbp]
  \centering 
  \includegraphics[width=.48\textwidth,scale=1.8]{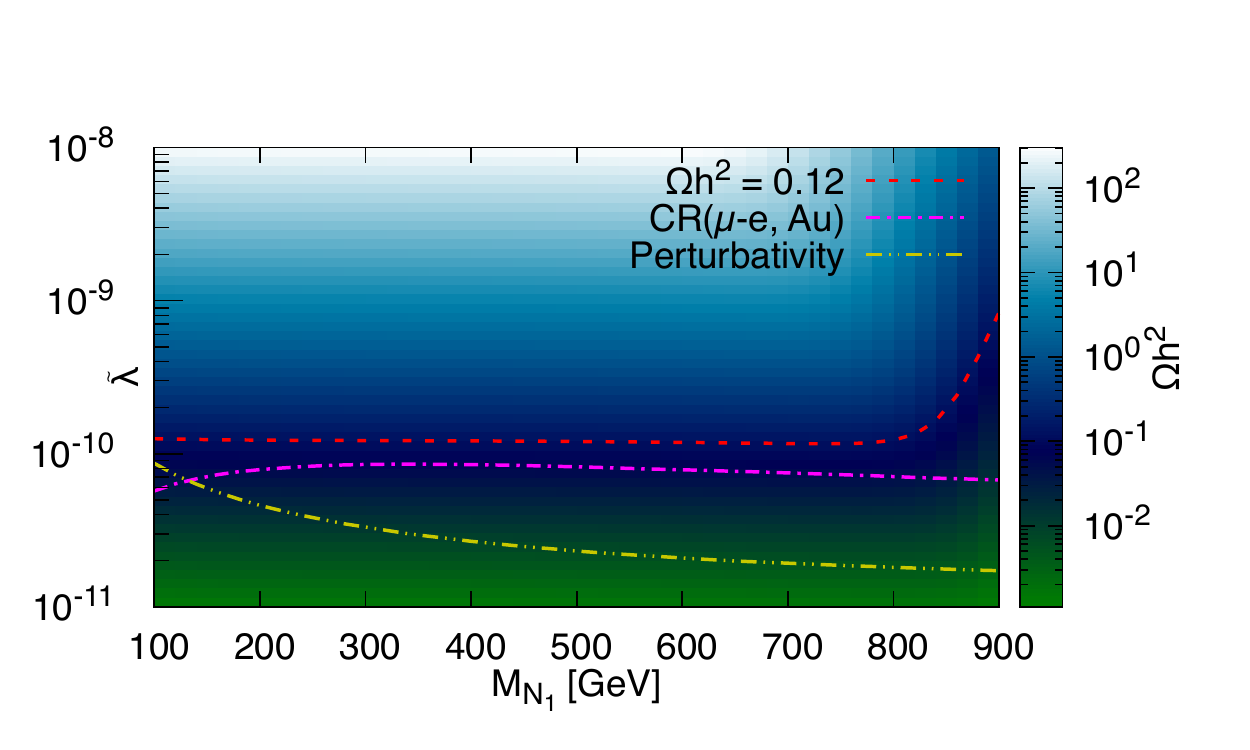}
  \includegraphics[width=.48\textwidth,scale=1.8]{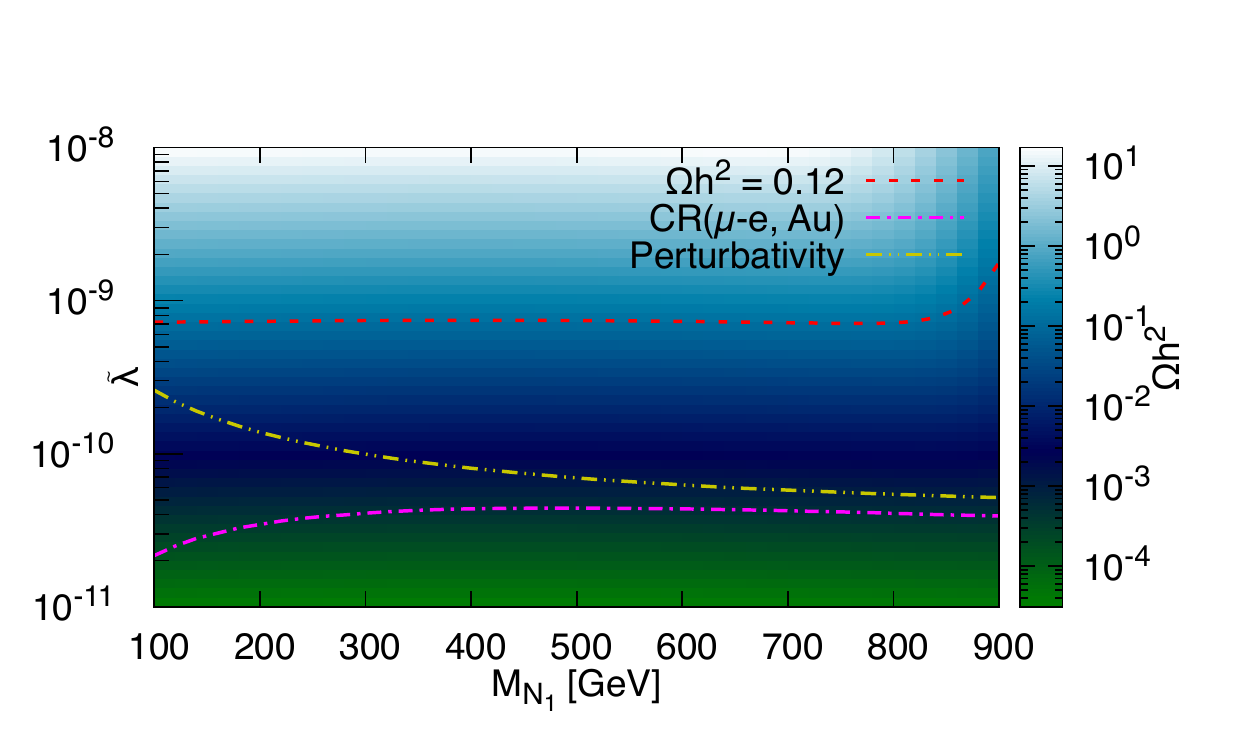}
  \includegraphics[width=.48\textwidth,scale=1.8]{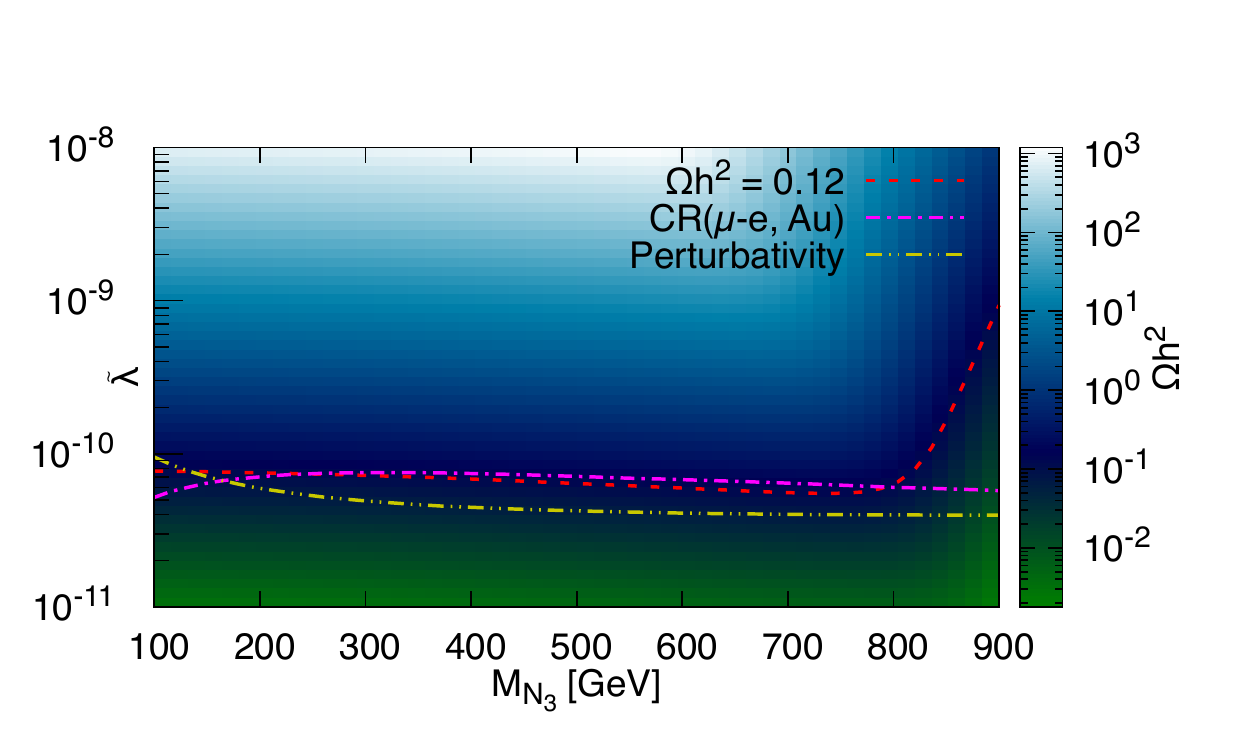}
  \includegraphics[width=.48\textwidth,scale=1.8]{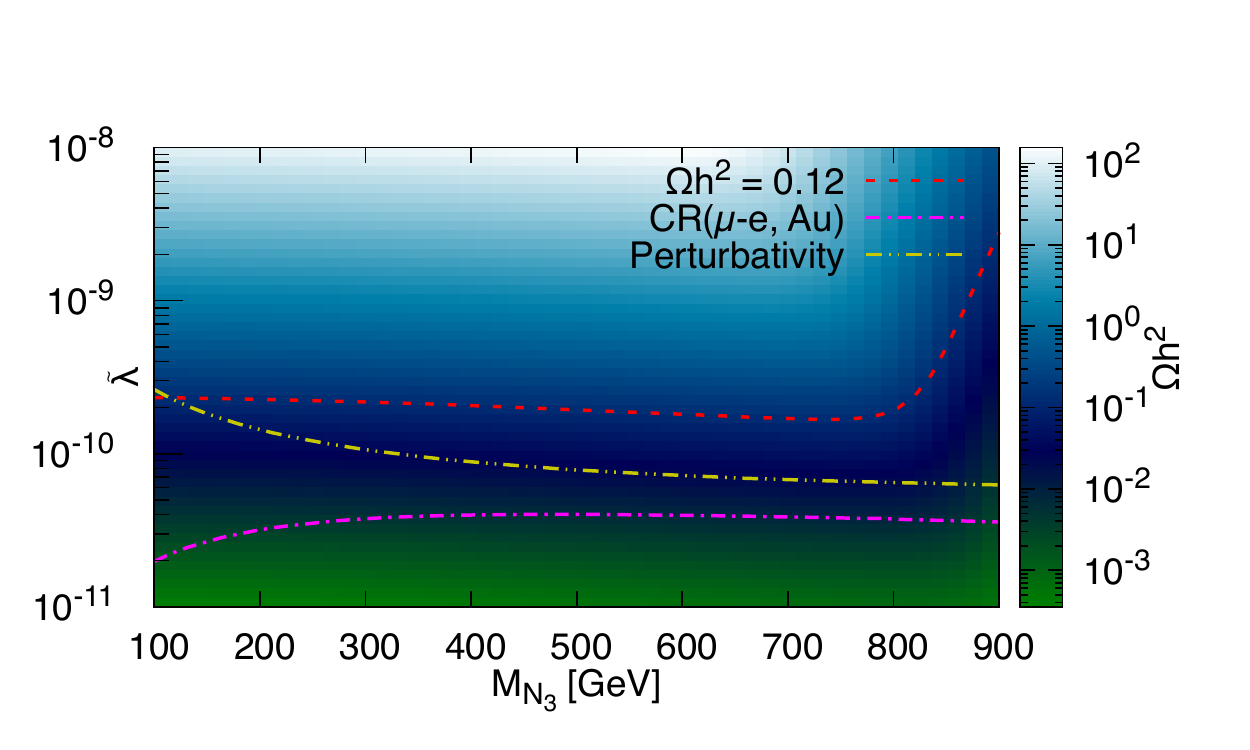}
  \caption{\label{fig:ex2}{The model prediction for the dark matter relic density (by the color gradient) is presented as a function of $M_{N_1}$ and $\widetilde{\lambda}$ for the normal mass hierarchy of the light neutrino masses and two fixed value of the lightest neutrino mass $m_1$: 0.02 GeV (top left panel) 0.06 GeV (top right panel). Similarly, the model prediction for the dark matter relic density (by the color gradient) is presented as a function of $M_{N_3}$ and $\widetilde{\lambda}$ for the inverted mass hierarchy of the light neutrino masses and two fixed value of the lightest neutrino mass $m_3$: 0.02 GeV (bottom left panel) 0.06 GeV (bottom right panel). Other details are same as in Fig.~\ref{fig:ex1}.}}
\end{figure*}

\section{\label{sec:DM} Constraints from Lepton Flavour Violation and Dark Matter relic density}
In this Section, we will discuss the constraints on the phenomenologically relevant parameters, namely, $\tilde \lambda$, $M_{N_1}$, and $m_1(m_3)$, resulting from the experimental upper bounds on the LFV observables and WMAP\cite{WMAP:2012fli} and Planck\cite{Planck:2015fie} measured value of the dark matter relic density. Due to $Z_2$ symmetry, the Scotogenic model includes several candidates for dark matter ranging from fermion type (the lightest $N_i$) to scalar type (the neutral components of the $Z_2$--odd scalar, namely $\phi_s$ and $\phi_p$). Dark matter searches have been carried out extensively for scalar and fermionic-type dark matter in the context of the Scotogenic model. The $Z_2$-odd scalar being doublet under $SU(2)_L$, the scalar dark matter scenarios suffer the strong constraints on dark matter nucleon scattering cross-section from the dark matter direct detection experiments. However, the singlet $Z_2$-odd fermions do not suffer direct detection constraints. Therefore, in this work, we focused on the scenario with a $Z_2$--odd singlet fermion $N_i$ ($N_1$ for NH and $N_3$ for IH) being the lightest and hence, a candidate for the dark matter. In the context of the Scotogenic model, scenarios with low mass fermionic DM and small Yukawa couplings (to circumvent the strong constraints from the charged LFV processes) have been studied, assuming freeze-in production of DM RD~\cite{}. In contrast, we studied a freeze-out scenario with relatively high mass (ranging from a few hundred GeVs to a TeV) dark matter and  large Yukawa couplings ${\cal{O}}(1)$ \footnote{The $Z_2$--odd fermions being singlet under the $SU(2)_L$, they can only interact with the SM matter fields via the Yukawa interactions. Therefore, large Yukawa allows them to be produced and studied at the collider experiments, making the scenario testable.}. To suppress the LFV constraints resulting from the large Yakawa couplings, we use the particular texture of the Yakawa matrix and the $Z_2$--odd fermion masses derived in section~\ref{sec:cLFV}. To explain the neutrino oscillation data and constraints from LFV observables for a given light neutrino mass hierarchy, the parametrization introduced in section~\ref{sec:cLFV} determines all the phenomenologically relevant parameters, namely the Yukawa matrix ($Y$) and the masses of the $Z_2$--odd singlet fermions ($M_{N_i}$), in terms of only three independent parameters \footnote{The parameters in the $Z_2$--odd scalar sector, namely $\mu_{\phi}$, $\lambda_1$ and $\lambda_2$, also remain independent.  However, these parameters do not significantly impact the phenomenology discussed in this work. Therefore,  we consider a few fixed benchmark values for those parameters to present the numerical results.}, namely $\tilde \lambda$, $M_{N_1}(M_{N_3})$, and $m_1(m_3)$ for NH(IH). 

\begin{figure*}[tbp]
  \centering 
  \includegraphics[width=.45\textwidth,scale=1.5]{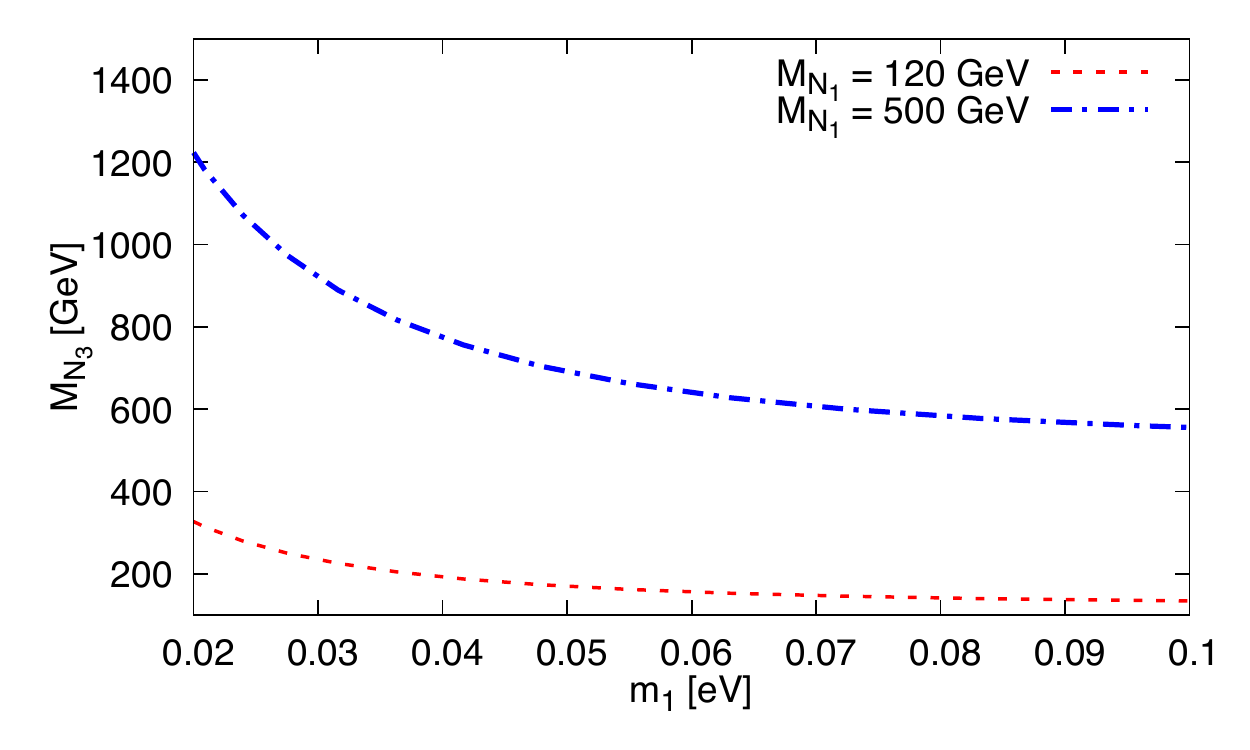}
  \hfill
  \includegraphics[width=.45\textwidth,scale=1.5]{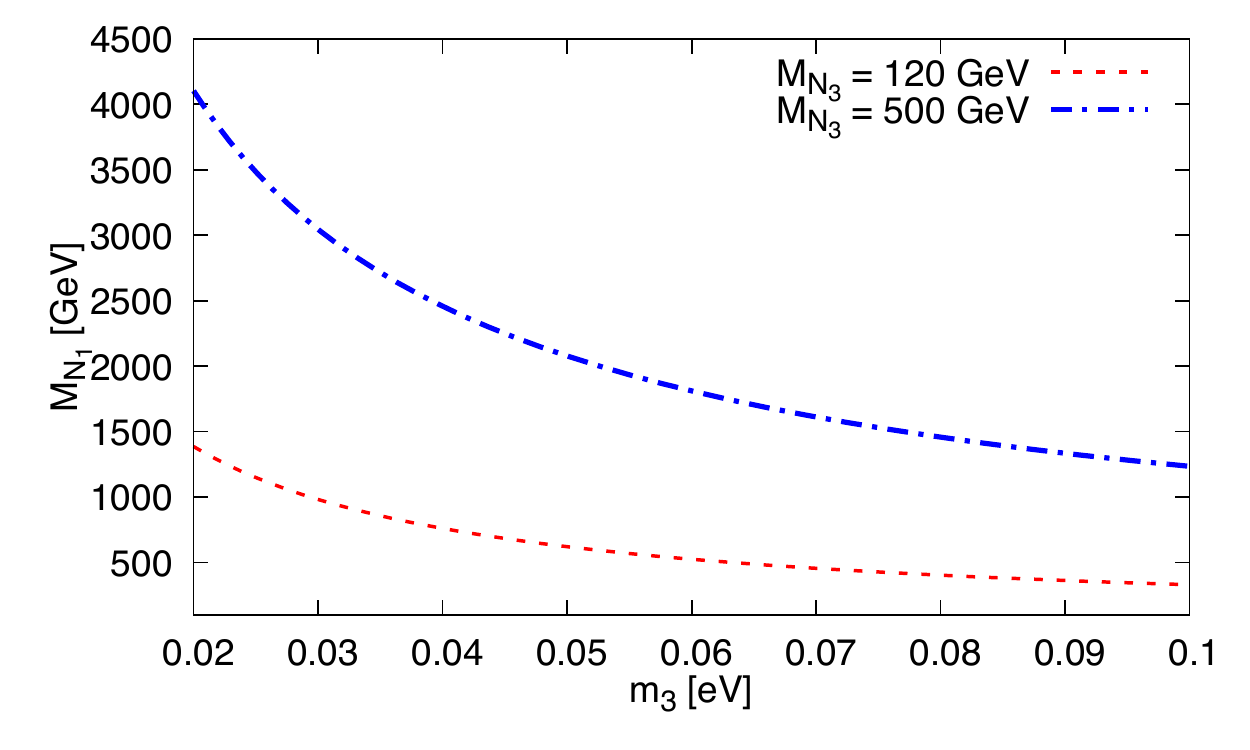}
  \caption{\label{fig:c1}Left panel: $M_{N_3}$ as a function of  $m_1$ in NH scenario for two fixed values of $M_{N_1}$. Right panel: $M_{N_1}$ as a function of $m_3$ in IH scenario for two fixed $M_{N_3}$.}
\end{figure*}

\begin{figure*}[tbp]
  \centering 
  \includegraphics[width=.48\textwidth,scale=1.5]{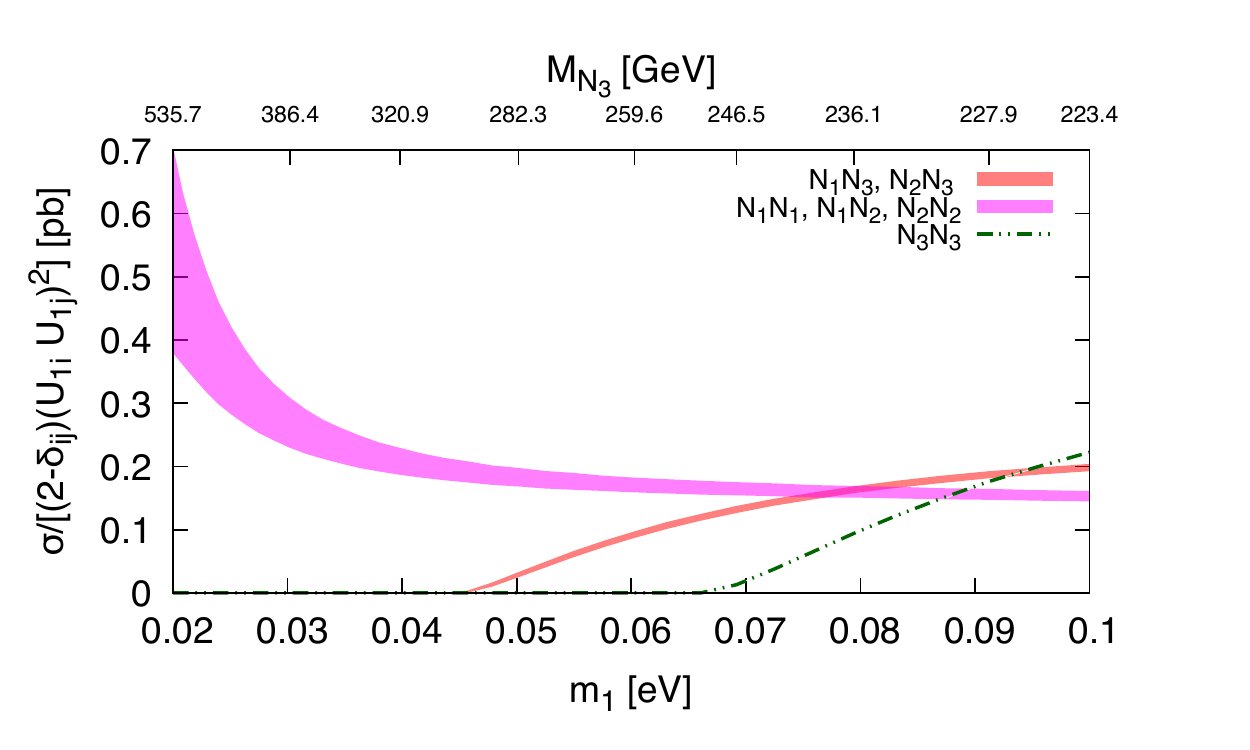}
  \includegraphics[width=.48\textwidth,scale=1.5]{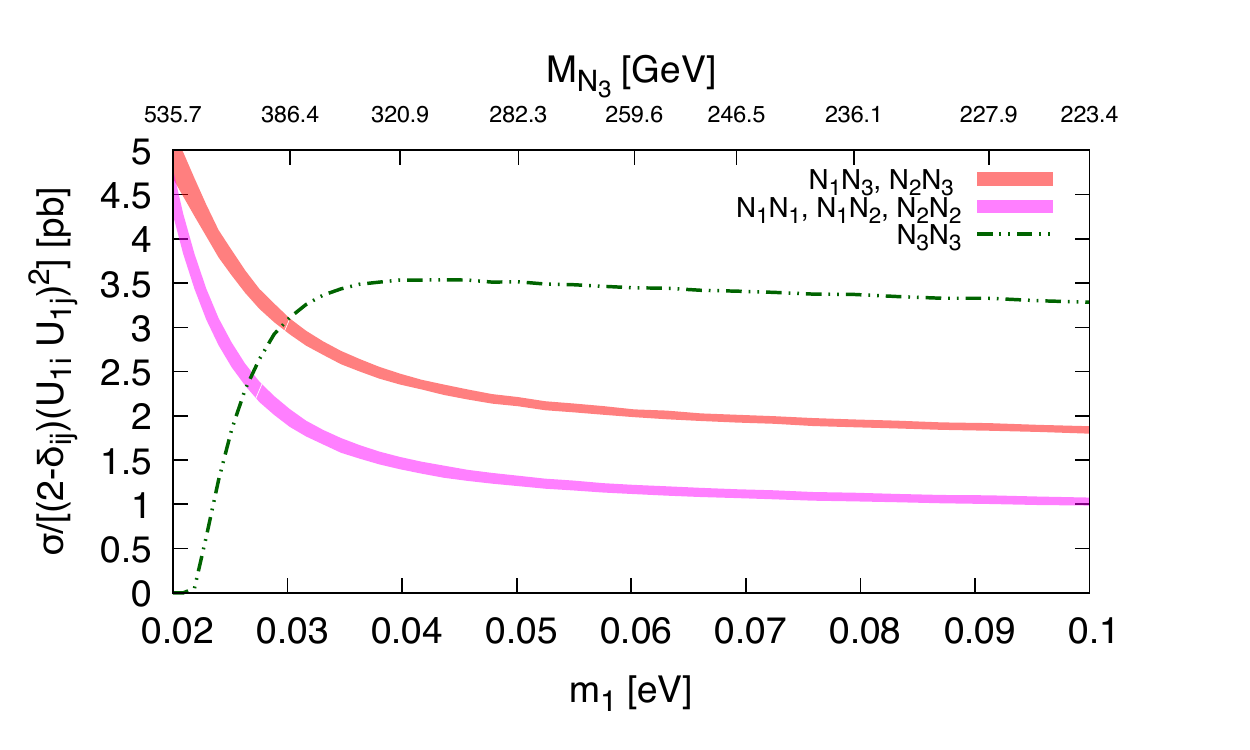}
  \caption{\label{fig:c2}Pair production cross-sections of different combinations of the $Z_2$--odd fermions are shown as a function of $m_1$ while keeping DM mass $M_{N_1}$ fixed at 200 GeV (NH scenario) at the electron-positron collider with centers of mass-energy 500 GeV (left) and 1 TeV (right).}
\end{figure*}

To present our numerical results, we have chosen three benchmark points (BP)s listed in table~\ref{table:A1}. Note that the first and the second row of the table~\ref{table:A1} contain all the independent parameters, whereas the parameters in the third row are determined from the parameterization introduced in section~\ref{sec:cLFV}. BP1 and BP2 correspond to the NH scenario with two different values for the dark matter mass. However, BP3 belongs to the IH scenario. The theoretical predictions for different LFV observables and the dark matter relic densities for the three benchmark points are presented in columns 3, 4, and 5 of table~\ref{table:A2}. The second column of table~\ref{table:A2} shows the experimental limits(measured value for the DM RD)  on those observables. Note that the first three rows of the table~\ref{table:A2} correspond to the branching ratios of three different LFV decays of the SM charged leptons into a lighter charged lepton in association with a photon (BR$l_\alpha \rightarrow  l_\beta$ $\gamma$). The theoretical predictions for these branching ratios are extremely suppressed. This is a consequence of the parameterization in section~\ref{sec:cLFV}, which has been designed to suppress such LFV decays. Although consistent with the experimental limits, other LFV observables are comparatively less suppressed, with some (in particular, CR($\mu - e, Al$)) being within the reach of future sensitivity of COMET \cite{Kuno:2013mha} and Mu2e \cite{Pezzullo:2017iqq} experiment. Although it is possible to obtain similar parameterizations to suppress another set of LFV observables, all the LFV observables can not be suppressed simultaneously. This makes some of the LFV observables sensitive to future experiments rendering testability to the model. 

To identify the experimentally consistent part of the parameter space, we performed a scanning over the independent parameters, namely $\tilde \lambda$, $M_{N_1}(M_{N_3})$, and $m_1(m_3)$ for NH(IH), keeping the other parameters in the $Z_2$-odd scalar sector fixed at some benchmark values. Figs.~\ref{fig:ex1} and \ref{fig:ex2} show the results of our parameter scan.  Fig.~\ref{fig:ex1} districts the model prediction for the dark matter relic density (by the color gradient) as a function of $m_1$ and $\widetilde{\lambda}$ for the normal mass hierarchy of the light neutrino masses and two fixed masses for the lightest $Z_2$-odd fermion (the dark matter candidate), namely  $M_{N_1}=$ 200 GeV (left panel) and 500 GeV (right panel). The values of the parameters in the $Z_2$-odd scalar sector are kept fixed at $\mu_{\phi}=1$ TeV, $\lambda_{1} = 0.004$ and $\lambda_{2}= 0.005$. For a given dark matter mass (in case of Fig.~\ref{fig:ex1}, 200 GeV for the left panel and 500 GeV for the right panel), the dark matter annihilation and hence, the relic density dominantly\footnote{The $Z_2$--odd scalars appear as t-channel propagator in the Feynman diagram of dark matter annihilating into the SM leptons. Therefore, the dark matter annihilation cross sections depend on the masses of these $Z_2$--odd fermions. However, the dependence is much weaker than the dependence on the Yukawa couplings.} depends on the Yukawa matrix ($Y$). Although the neutrino oscillation data and bounds from the LFV experiments completely determine the texture of the Yukawa matrix ($Y$), the strength of the Yukawa couplings (see Eq.~\ref{eq:b7}) depends on the free parameters like $m_1$, $\widetilde{\lambda}$ e.t.c. For example, the Yukawa matrix ($Y$) in the Yukawa matrix ($Y$) is inversely proportional to the square root of $\widetilde{\lambda}$. Therefore, smaller $\widetilde{\lambda}$ corresponds to larger Yukawa couplings and hence, large dark matter annihilation in the early universe and small relic density. The dependence of the Yukawa couplings on the lightest SM neutrino mass is a little complicated and is not directly visible in Eq.~\ref{eq:b7}. On one hand, the lightest neutrino mass determines the masses of the other SM neutrinos which appear in the matrix $\tilde M_\nu$; on the other hand, the heavy $Z_2$--odd neutrino masses also depend on the lightest SM neutrino mass because of the parameterization introduced in Eq.~\ref{req4}. Therefore, in Eq.~\ref{eq:b7}, the dependence on $m_1$ appears in both $\tilde M_\nu$ and ${\cal M}$.  Smaller $m_1$ yields smaller Yukawa couplings and hence, larger relic density. The red dashed line in Fig.~\ref{fig:ex1} shows the points on the parameter space consistent with the measured value of the relic density. Points above the red dashed line result in relic densities larger than the measured value and hence, ruled out. After suppressing the most constraining LFV decay, $\mu \to e \gamma$, using the parametrization in section~\ref{sec:cLFV}, the next constraining LFV bound result from the upper limit on CR$(\mu-e, Au) = 7.0 \times 10^{-13}$. The violet dash-dotted corresponds to CR$(\mu-e, Au) = 7.0 \times 10^{-13}$. Because of the larger Yukawa couplings, the region below this line yields CR$(\mu-e, Au) > 7.0 \times 10^{-13}$ and hence, ruled out. The yellow dash-dotted line shows the perturbativity bound below which some of the Yukawa couplings become non-perturbative ($> 3$). Fig.~\ref{fig:ex2} shows the model prediction for the dark matter relic density (by the color gradient) as a function of $M_{N_1}$ and $\widetilde{\lambda}$ for the normal mass hierarchy of the light neutrino masses and two fixed value of the lightest neutrino mass $m_1$: 0.02 GeV (left panel) 0.06 GeV (right panel). Fig.~\ref{fig:ex2} shows that the dark matter relic is relatively insensitive on the dark matter mass over a large range. The sudden decrease in the dark matter relic density for large $M_{N_1}$ is a consequence of the fact that dark matter co-annihilation with the $Z_2$--odd scalars starts dominating in those regions because of the particular choice for the scalar parameters: $\mu_{\phi}=1$ TeV, $\lambda_{1} = 0.004$ and $\lambda_{2}= 0.005$. The important point to be noted is that the allowed parameter space region is the region enclosed by the three lines corresponding to three different bounds. However, only the points on the red lines explain the measured value of the dark matter relic density, i.e., for given values of the lightest SM neutrino and the lightest $Z_2$--odd fermion mass, the value of $\widetilde{\lambda}$ gets fixed to obtain the observed relic density. Therefore, if we demand the model to explain the WMAP \cite{WMAP:2012fli} and Planck \cite{Planck:2015fie} measured value of the dark matter relic density, one of the three independent parameters gets determined in terms of the other two. This further reduces the number of independent parameters and makes the scenario more predictive at the collider experiments. The phenomenology of exotic fermions at the colliders which will be discussed in the following, is determined by two parameters, namely the lightest SM neutrino and the lightest $Z_2$--odd fermion mass.


\section{\label{sec:Coll}Collider Signatures}

In the presence of a set of exotic fermions and scalars at TeV scale, it is instructive to study the signatures of these exotics at the collider experiments. However, the $Z_2$ symmetry forbids single production of these exotics. The $Z_2$-odd scalars being doublet under the $SU(2)_L$ have gange interactions with the SM gauge bosons; hence, the $Z_2$-odd scalars can be pair produced at the LHC via Drell-Yan (DY) processes. However, the $Z_2$-odd fermions being singlet under the SM gange symmetry do not have any gange interactions or Yukawa interactions with the quarks. Therefore, the $Z_2$-odd fermions can not be produced directly at the LHC. At the LHC, they can arise from the decay of $Z_2$-odd scalars. However, due to the large Yukawa couplings with the SM leptons, the $Z_2$-odd fermions can be copiously pair-produced at future electron-positron collider experiments. In this work, we primarily focused on the signatures of the $Z_2$-odd fermions at the electron-positron colliders. Before going into the details of collider signatures, it is important to understand the mass spectrums of the $Z_2$-odd fermions, their production, and decay which will be discussed in the following. 

Due to the parameterization introduced in section~\ref{sec:cLFV}, two parameters completely determine the masses of the $Z_2$-odd fermions: the masses of the lightest SM neutrino and the lightest $Z_2$-odd fermion (see Eq.~\ref{req4}).  In Fig.~\ref{fig:c1}, we have shown the mass of the $N_3$ (left panel) and $N_1$ (right panel) as a function of the mass of the lightest SM neutrino mass ($m_1$ for NH and $m_3$ for IH), respectively for two fixed values of the dark matter mass. In both scenarios, the mass of $N_2$ is almost degenerate with the mass of the lightest $Z_2$-odd fermions (with $M_{N_1}$ in NH and  $M_{N_3}$ in IH).

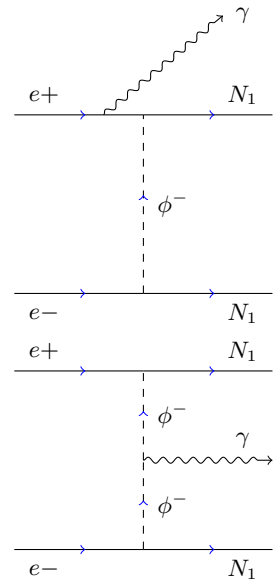
\begin{figure}
  \begin{center}
    \begin{tikzpicture}[line width=1. pt, scale=1.32,every node/.style={scale=1.0}]
      \draw[fermion,black,thin] (-1.3,0) --(0,0);
      \draw[fermion,black,thin] (0,0) --(1.3,0.0);

      \draw[fermion,black,thin] (-1.3,1.8) --(0,1.8);
      \draw[fermion,black,thin] (0.0,1.80) --(1.3,1.8);
      \draw[scalar,black,thin] (0.0,0) --(0,1.8);
      
      \draw[vector,black,thin] (-0.4,1.8) -- (0.8,2.8);   

      \node at (-1.0,-0.2) {$e-$};
      \node at (1.0,-0.2) {$N_1$};

      \node at (-1.0,2.0) {$e+$};
      \node at (1.0,2.0) {$N_1$};
      \node at (1.0,2.8) {$\gamma$};
      \node at (0.30,0.9) {$\phi^-$};
    \end{tikzpicture}

    \begin{tikzpicture}[line width=1. pt, scale=1.32,every node/.style={scale=1.}]
      \draw[fermion,black,thin] (-1.3,0) --(0,0);
      \draw[fermion,black,thin] (0,0) --(1.3,0.0);

      \draw[fermion,black,thin] (-1.3,1.8) --(0,1.8);
      \draw[fermion,black,thin] (0.0,1.80) --(1.3,1.8);
      \draw[scalar,black,thin] (0.0,0) --(0,0.9);
      \draw[scalar,black,thin] (0.0,0.9) --(0,1.8);
      \draw[vector,black,thin] (0.0,0.9) -- (1.3,0.9);   

      \node at (-1.0,-0.2) {$e-$};
      \node at (1.0,-0.2) {$N_1$};

      \node at (-1.0,2.0) {$e+$};
      \node at (1.0,2.0) {$N_1$};
      \node at (1.0,1.1) {$\gamma$};
      \node at (0.30,0.45) {$\phi^-$};
      \node at (0.30,1.35) {$\phi^-$};
    \end{tikzpicture}

  \end{center}
  \caption{\label{fig:cs1}Feynman diagram of mono-photon + $\slashed{E}_T$ channel.}
\end{figure}
\begin{figure*}[tbp]
  \centering
  \includegraphics[width=.48\textwidth,scale=1.0]{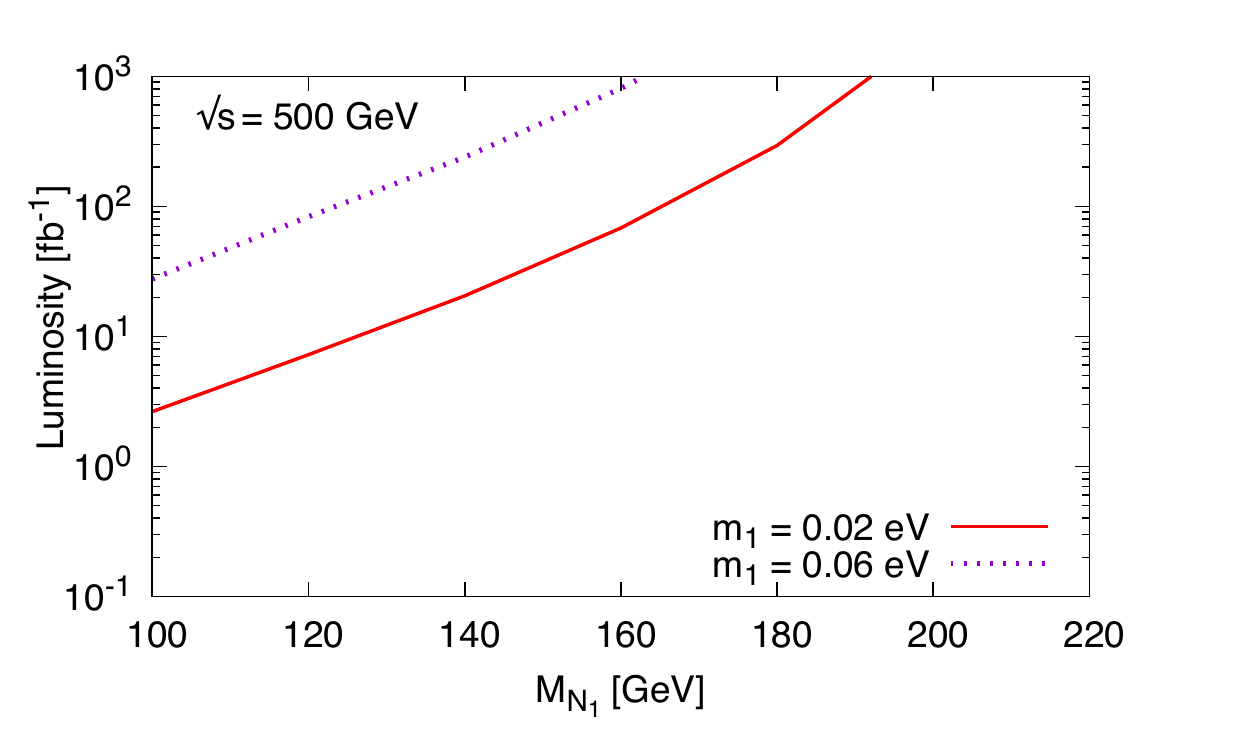}
  \hfill
  \includegraphics[width=.48\textwidth,scale=1.0]{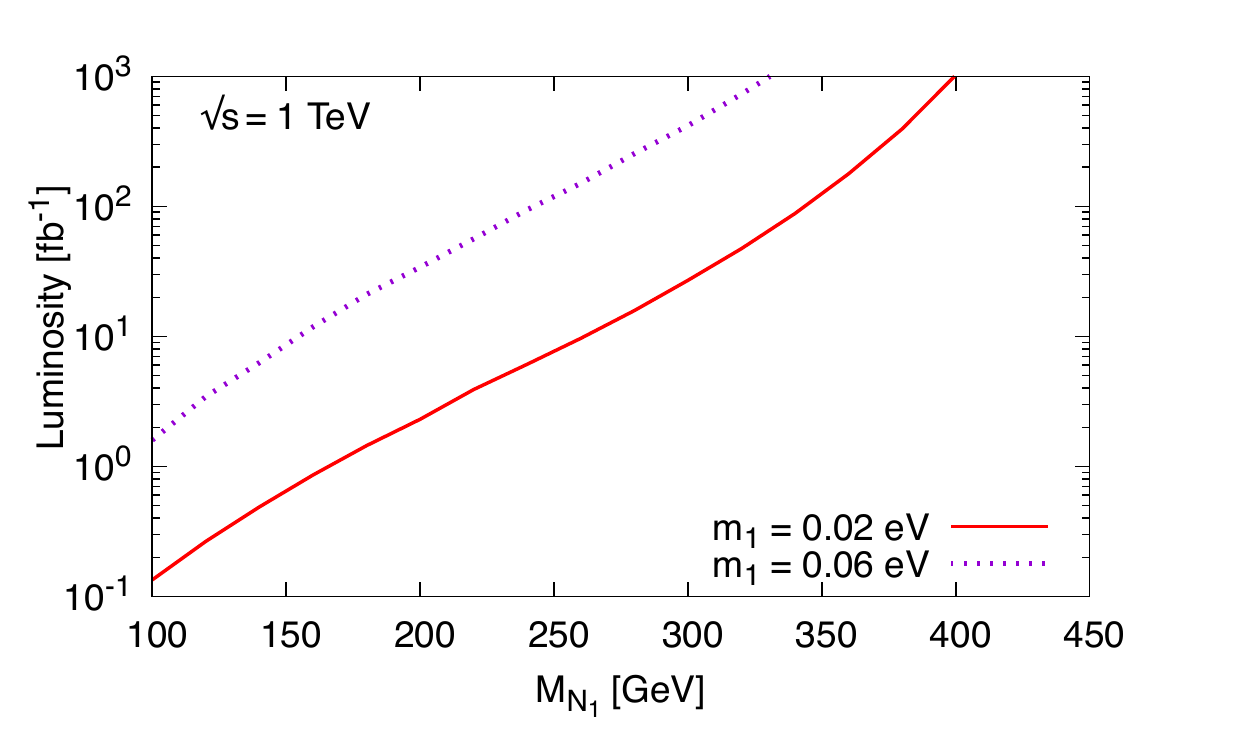}
  \caption{\label{fig:cs2}The required luminosities as a function of $M_{N_1}$ for 5$\sigma$ discovery in mono-photon + $\slashed{E}_T$ channel at the center of mass energy 500 GeV (left panel) and 1 TeV (right panel).}
\end{figure*}

\subsection{\label{subsec:Prod}Production and Decay of the exotic fermions}
The exotic fermions can be pair produced at an electron-positron collider via the Yukawa interactions involving the electron, the  $Z_2$--odd scalar, and fermion. The pair production proceeds through a t-channel diagram involving the exotic scalar in the t-channel. In Fig.~\ref{fig:c2}, we show the pair production cross-sections of all possible combinations of the exotic fermions as a function of the lightest neutrino mass $m_1$ in NH scenario at the electron-positron collider with  the center of mass energy  500 GeV (left panel ) and 1 TeV (right panel) for $M_{N_1}=200$ GeV. Note that the lightest SM and $Z_2$--odd neutrino masses  completely determine the collider phenomenology of the scenario since $\widetilde{\lambda}$ gets fixed to explain the relic density data (see the discussion in section \ref{sec:cLFV}). $N_1$ and $N_2$ being almost mass degenerate for NH scenario, in Fig.~\ref{fig:c2}, we grouped different pair production channels according to the similarity in the kinematics. For example, the kinematics of associated productions of $N_3$ in association with $N_1$ or $N_2$ are similar and hence, can be grouped together. However, the Yukawa interactions involved in $\sigma(N_1 N_3)$ and $\sigma(N_2 N_3)$ are different. While the former is proportional to $\left(Y^{11}Y^{13}\right)^2$, the later is proportional to $\left(Y^{12}Y^{13}\right)^2$ which leads to $\frac{\sigma(N_1 N_3)}{\left(U_{PMNS}^{11}U_{PMNS}^{13}\right)^2} \approx \frac{\sigma(N_1 N_3)}{\left(U_{PMNS}^{12}U_{PMNS}^{13}\right)^2}$ for the Yukawa matrix $Y$ given in Eq.~\ref{eq:b7}. The bands in Fig.~\ref{fig:c2} correspond to $U_{PMNS}$ scaled production cross-sections for three distinct sets of pair-productions that can be grouped together. The small mass splitting between $N_1$ and $N_2$ is responsible for the band instead of a line.

\begin{figure}
  \begin{center}
    \begin{tikzpicture}[line width=1.4 pt, scale=1.32,every node/.style={scale=1.0}]
      \draw[fermion,black,thin] (-1.3,0) --(0,0);
      \draw[fermion,black,thin] (0,0) --(1.3,0.0);

      \draw[fermion,black,thin] (-1.3,1.8) --(0,1.8);
      \draw[fermion,black,thin] (0.0,1.80) --(1.3,1.8);
      \draw[scalar,black,thin] (0.0,0) --(0,1.8);
      
      \draw[fermion,black,thin] (1.3,0.0) --(1.8,0.6);
      \draw[scalar,black,thin] (1.3,0.0) --(1.8,-0.5);

      \draw[fermion,black,thin] (1.8,-0.5) --(2.2,-0.3);
      \draw[fermion,black,thin] (1.8,-0.5) --(2.2,-1.0);

      \node at (-1.0,-0.2) {$e-$};
      \node at (1.0,-0.2) {$N_3$};

      \node at (-1.0,2.0) {$e+$};
      \node at (1.0,2.0) {$N_1$};
      \node at (0.30,0.9) {$\phi^-$};

      \node at (2.,0.6) {$l_i$};
      \node at (2.0,-.06) {$ \phi^\pm,\phi_{s,p}$};

      \node at (2.4,-0.4) {$l_j$};
      \node at (2.4,-1.2) {$N_{1,2}$};
    \end{tikzpicture}
  \end{center}
  \caption{\label{fig:cs3}Feynman diagram of OSD $+ \slashed{E}_T$ channel.}
\end{figure}
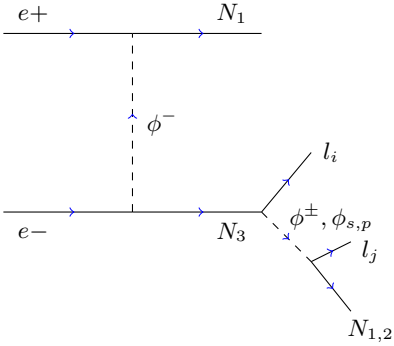

After being produced at the colliders, the $Z_2$--odd exotics decays into lighter $Z_2$--odd exotics associated with one or more SM particles. In absence of any Kinematically allowed decay mode, the lightest $Z_2$--odd exotic remains stable and, being weakly interacting, remains invisible at the detectors resulting in missing energy signature. Therefore, the pair production of $Z_2$--odd fermions leads to multiple leptons/jets in association with large missing energy signatures at the electron-positron collider. Before going into the collider signatures, a brief discussion about the possible decay modes of the $Z_2$--odd exotics is given in the following.

In NH scenario, $N_{1}$ being the lightest $Z_2$--odd exotics remains stable. In the scenario with the $Z_2$--odd scalars being heavier than the $Z_2$--odd fermions, $N_{2,3}$ undergoes tree-level 3-body decays into a pair of SM leptons in association with a lighter $Z_2$--odd fermion. These 3-body decays are mediated by off-shell exotic scalars ($\phi^+,\phi_{s/p}$) via the Yukawa interactions involving the SM lepton doublet, $Z_2$--odd singlet fermions and doublet scalar. The 3-body decay width  $\Gamma(N_{2,3} \rightarrow \overline{l_\alpha} l_\beta N_1)$ is given by,
\begin{eqnarray}
  \hspace{-25mm}
  \Gamma(N_{2,3} \rightarrow \overline{l_\alpha} l_\beta N_1) = \frac{M^5_{N_{2,3}}}{6144 \pi^3 m^4_S} (|Y^{\beta 1}|^2|Y^{\alpha 2,3}|^2 \nonumber\\
  ~~~~~~~~~~~~~~~~~~~~~~~~~~~~~~~~~+ |Y^{\alpha 1}|^2|Y^{\beta 2,3}|^2)
  \label{D1}
\end{eqnarray}
, where $m_S$ is the mass of the mediating scalar. The $Z_2$--odd scalars being heavier than the $Z_2$--odd fermions, undergoes two-body decays into $Z_2$--odd fermions in association with a SM charged lepton or neutrino. The decay widths for the charged and the neutral $Z_2$--odd scalars are given by, 
\begin{eqnarray}
 \hspace{-10mm} \Gamma(\phi_{s/p}\rightarrow  \nu_\alpha N_\beta) = \frac{m_{\phi_{s/p}}|Y^{\alpha \beta}|^2 }{32\pi}\left(1-\frac{M^2_{N_\beta}}{m^2_{\phi_{s/p}}}\right)^2
  \label{D2}
\end{eqnarray}
\begin{eqnarray}
 \hspace{-10mm} \Gamma(\phi^+\rightarrow \overline{l_\alpha} N_\beta) = \frac{m_{\phi^+}|Y^{\alpha \beta}|^2 }{16\pi}\left(1-\frac{M^2_{N_\beta}}{m^2_{\phi^+}}\right)^2
  \label{D3}
\end{eqnarray}

\subsection{\label{subsec:Sig}Collider Signatures}
Being the lightest $Z_2$--odd particles, the largest contribution to the pair productions of $Z_2$--odd fermions at the electron-positron collider result from the pair and associated production of $N_1$ and $N_2$ in NH scenario (see Fig.~\ref{fig:c2}). $N_1$, being stable and weakly interacting, remains invisible in the detectors. Although having the most significant contribution to the pair productions, the $N_1 N_1$-production leads to the invisible final state at the electron-positron collider. Therefore, we consider the pair production of $N_1$ in association with a hard photon ($e^+ e^- \to N_1 N_1 + \gamma$) and study {\em mono-photon plus large missing energy final state} as a signature of $N_1$ at the electron-positron collider. Although the pair and associated production of $N_3$ are suppressed due to its large mass, the decay of $N_3$ into a pair of SM charged lepton and missing energy gives rise to interesting final states at the collider experiments. In this work, we consider the associated production of $N_3$ and study {\em the opposite sign di-lepton (OSD) plus missing energy final states} at the electron-positron collider.

\noindent {\em \bf 1. Mono-photon plus missing energy signature:} The pair production of $N_1$ at $e^+-e^-$ colliders along with a photon gives rise to mono-photon in association with large missing energy $\slashed{E}_T$ final state. Feynman diagram associated with this signal is depicted in Fig.~\ref{fig:cs1}. The dominant SM background contribution results from the production of neutrino anti-neutrino pairs in association with a photon. The signal and background events are generated in $MadGraph5-2.6.7$ \cite{Alwall:2014hca} with the initial state radiation (ISR) being simulated using the plugin provided in "Initial State Radiation Simulation with MadGraph" \cite{Li:2018qnh}. We define, the signal will discover with more than $S\sigma$ significance for a integrated luminosity $\mathcal{L}$ if
\begin{eqnarray}
  \frac{N_S}{\sqrt{N_B}} \geq S
  \label{D4}
\end{eqnarray}
where, $N_{S(B)}=\sigma_{S(B)}\mathcal{L}$ is the signal (background) events with integrated luminosity $\mathcal{L}$ and signal (background) cross-section $\sigma_{S(B)}$. The required luminosity for 5$\sigma$ discovery significance for a fixed lightest SM neutrino mass has been shown in Fig.~\ref{fig:cs2} as a function of $N_1$ mass. In Fig.~\ref{fig:cs2}, red (solid) and blue (dotted) lines correspond to the required luminosity for 5$\sigma$ discovery for the lightest SM neutrino mass 0.02 eV and 0.06 eV, respectively. The Left and right panel corresponds to 500 GeV and  1 TeV center of mass energy of the electron-positron collider.

\begin{table}[ht!]
  \centering
  \scalebox{.75}{\begin{tabular}{|c|c|}\hline
      BP1 (NH) &BP2 (NH)\\\hline
      \multicolumn{2}{|c|}{$\mu_\phi$ = 1 TeV, $\lambda_1$ = 0.004, $\lambda_2$ = 0.005}\\\hline
      $\widetilde{\lambda}$ = $1.24 \times 10^{-10}$, $M_{N_1} = 120$ GeV
      &$\widetilde{\lambda}$ = $7.32 \times 10^{-10}$, $M_{N_1} = 200$ GeV\\
      $m_1$ = $2.0 \times 10^{-11}$ GeV
      &$m_1$ = $6.0 \times 10^{-11}$ GeV\\\hline
      $M_{N_2}$ = 130.07 GeV, $M_{N_3}$ = 326.02 GeV  & $M_{N_2}$ = 201.94 GeV, $M_{N_3}$ = 260.08 GeV\\

      $Y$=$\begin{pmatrix}
        2.29 &-1.29 &0.67\\
        1.40 &1.64 &-1.65\\ 
        0.41 &1.84 &2.17\\
      \end{pmatrix}$
      &$Y$ = $\begin{pmatrix}
        1.30 &-0.73 &0.38\\
        0.79 &0.93 &-0.94\\
        0.22 &1 &1.18\\
      \end{pmatrix}$\\\hline
  \end{tabular}}
  \caption{\label{table:a1}Benchmark points for signal OSD $+ \slashed{E}_T$.}
\end{table}

\begin{figure*}[tbp]
  \centering 
  \includegraphics[width=.24\textwidth,scale=.50]{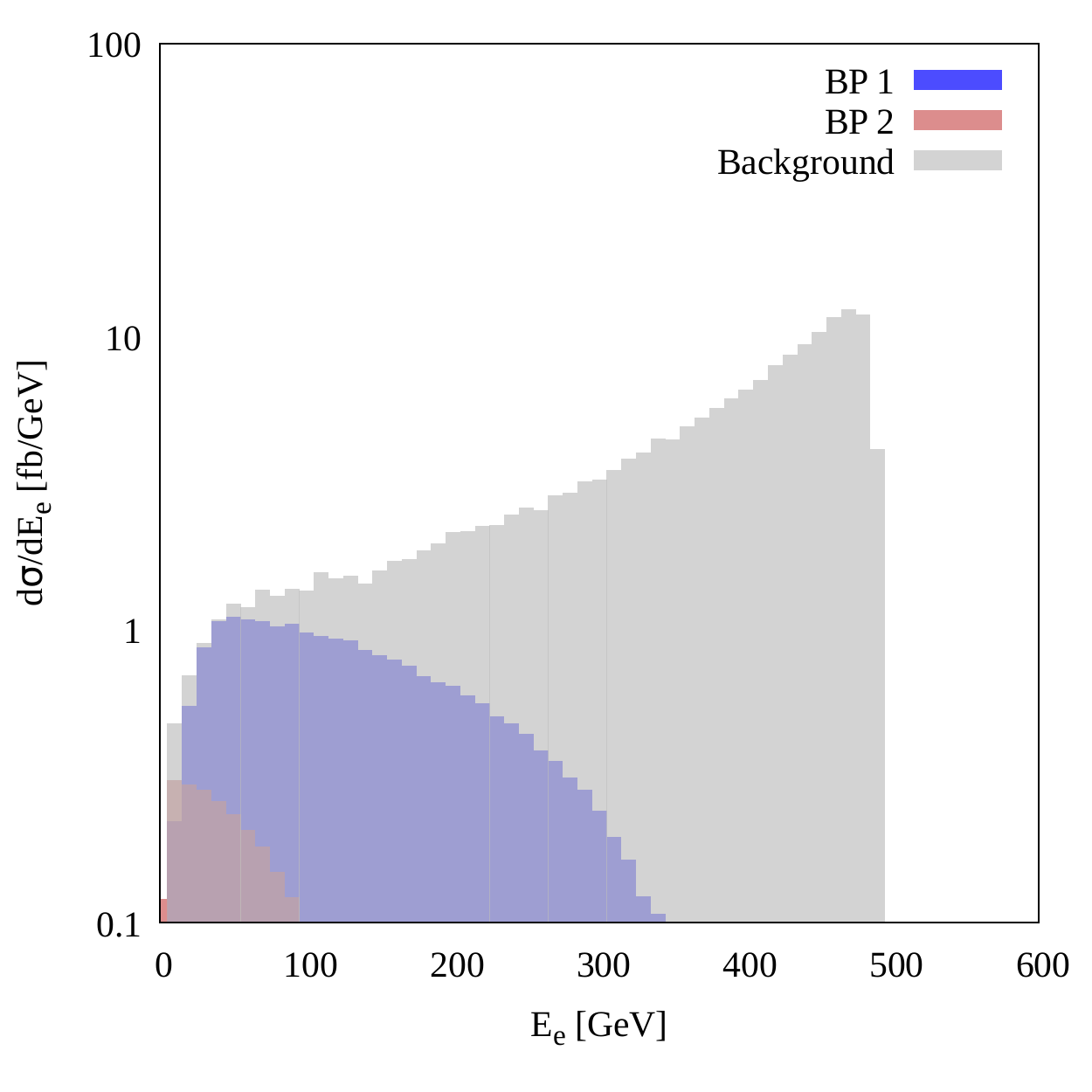}
  \hfill
  \includegraphics[width=.24\textwidth,scale=.50]{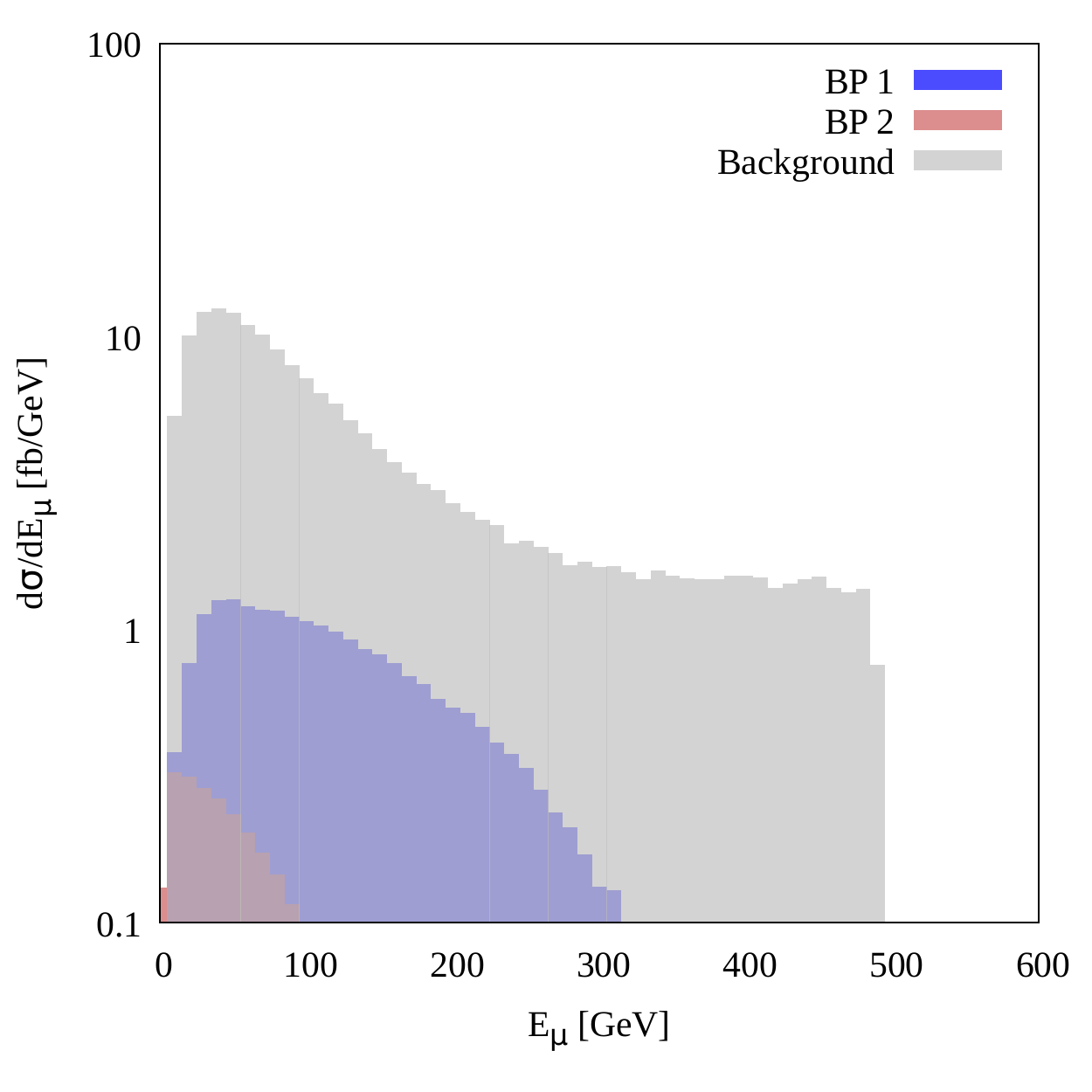}
  \hfill
  \includegraphics[width=.24\textwidth,scale=.50]{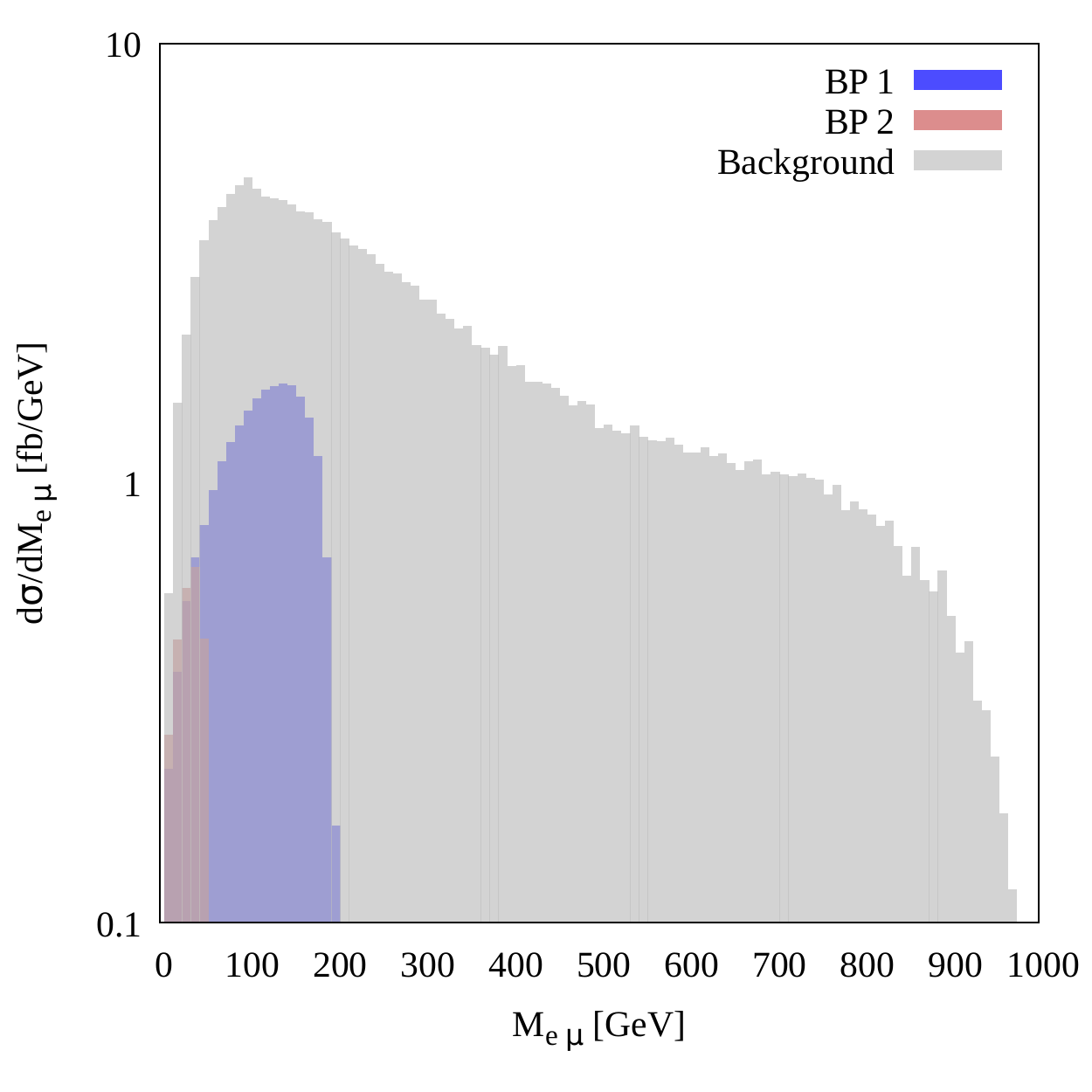}
  \hfill
  \includegraphics[width=.24\textwidth,scale=.50]{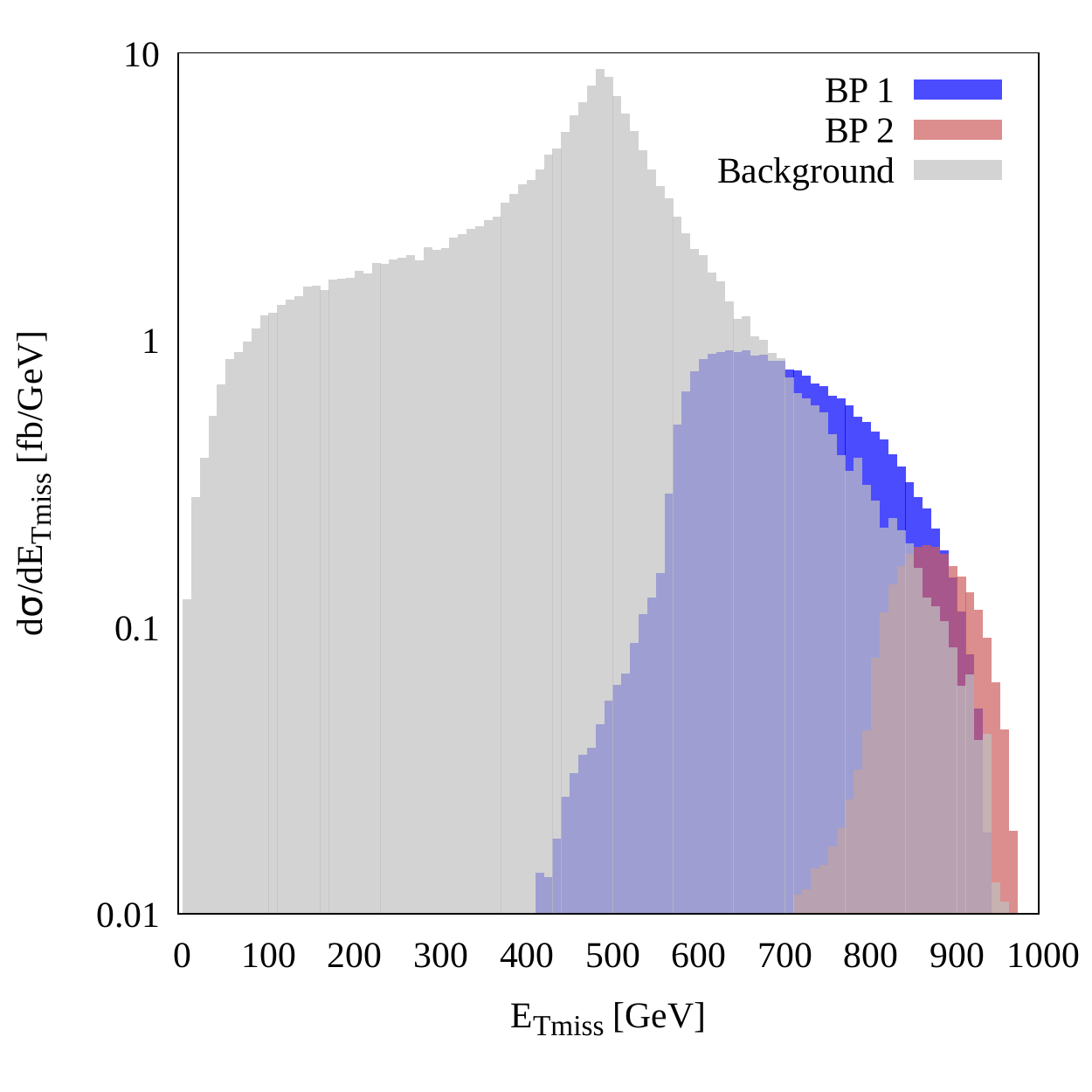}
  \caption{\label{fig:a2} The characteristic distributions regarding signal $e^\pm \mu^\mp +\slashed{E}_T$ for BPs and the SM background.}
\end{figure*}

\begin{figure*}[tbp]
  \centering 
  \includegraphics[width=.48\textwidth,scale=.50]{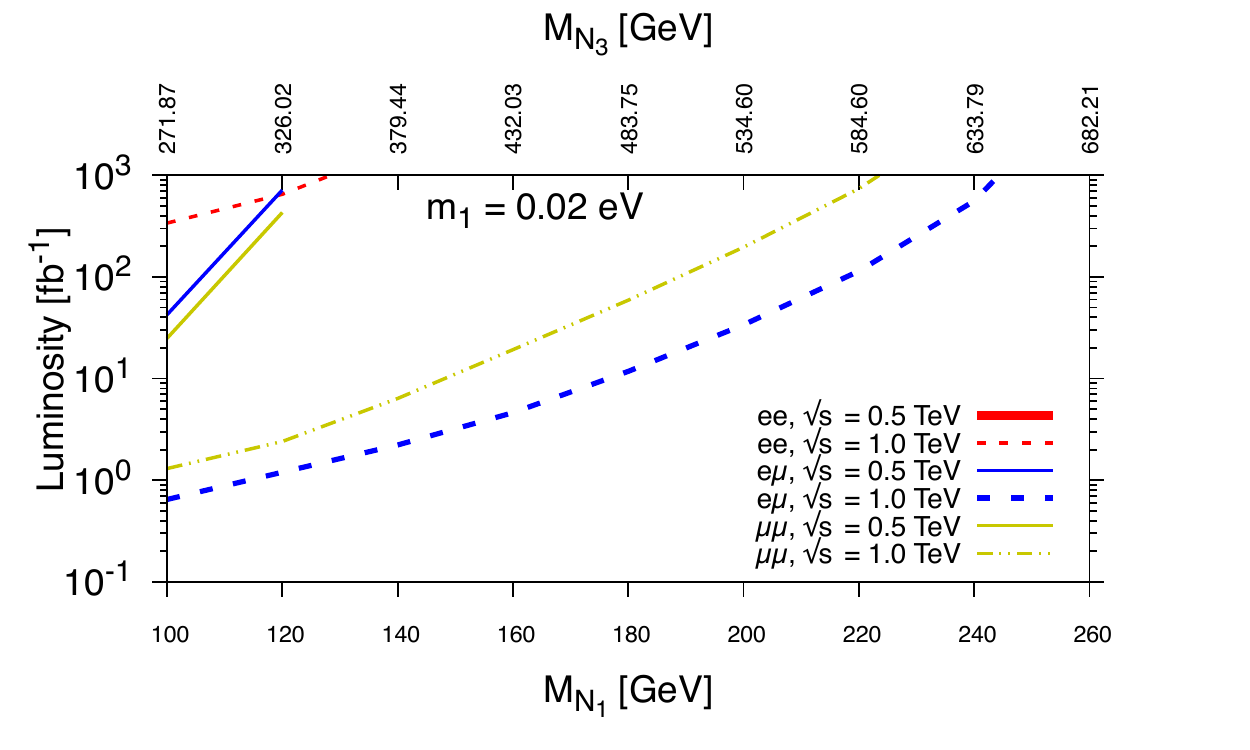}
  \hfill
  \includegraphics[width=.48\textwidth,scale=.50]{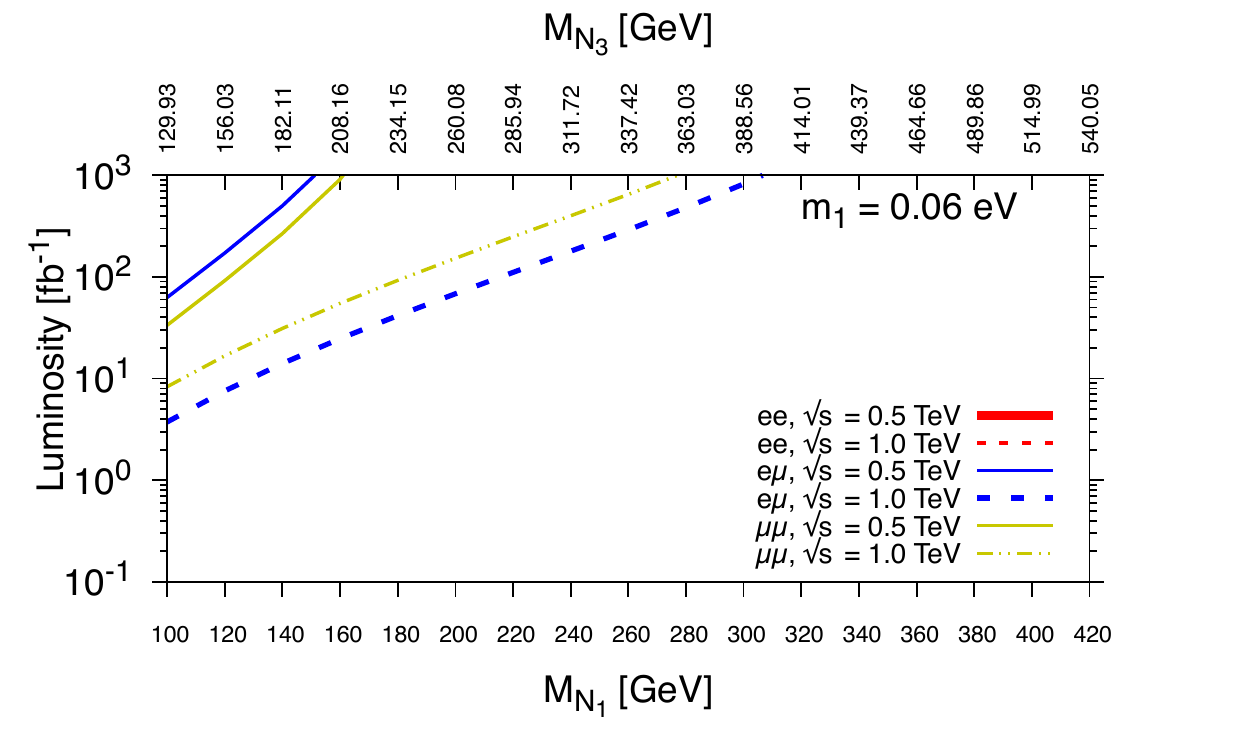}
  \caption{\label{fig:a3} The required luminosities as a function o $M_{N_1}$ for 5$\sigma$ discovery in different OSD $+\slashed{E}_T$ channels at center of mass energy 500 GeV and 1 TeV for the fixed lightest neutrino mass 0.02 eV (left panel) and 0.06 eV (right panel).}
\end{figure*}

{\em \bf 2.  Opposite sign di-lepton plus missing energy signature}: Fig.~\ref{fig:cs3} depicts the Feynman diagram leading to opposite sign di-leptons (OSD) and large missing energy $\slashed{E}_T$ signatures. To present numerical results, we have taken two BPs as defined in Table \ref{table:a1}. BP1 and BP2 belong to NH scenario from Fig. \ref{fig:ex2} for the lightest SM neutrino masses 0.02 eV and 0.06 eV respectively. Dominant SM backgrounds result from the production of $l^\pm l^\mp \nu_l \bar{\nu}_l$, $t \bar{t} ~{\rm and}~ \tau \bar{\tau}$. The signal and backgrounds are generated using $MadGraph5-2.6.7$ \cite{Alwall:2014hca} with the following sets of acceptance cuts: min ${p}_{T{e,\mu}} =$ 10 GeV, $|\eta_{e,\mu}| \leq 2.5$, min $\Delta R_{e\mu}=$ 0.4.  The characteristic kinematic distributions for the signal BPs and the SM background are depicted in Fig. \ref{fig:a2} for $E^\pm \mu^\mp + \slashed{E}_T$ channel. In view of these signals and the background distributions, the proposed event selection criteria are tabulated in Table \ref{table:a3}. The efficiency of our proposed cuts is demonstrated through cut flow for signal $e^\pm\mu^\mp+\slashed{E}_T$ at $\sqrt{s}=1$ TeV in Table \ref{table:a4}. Cross-sections for both BPs and the SM background are shown in femtobarns (fb). Fig. \ref{fig:a3} shows the required luminosity for $5\sigma$ discovery of OSD $+\slashed{E}_T$ signal over the SM backgrounds as a function of $N_1$ mass for two fixed values of the lightest neutrino mass and $\sqrt{s}=$ 500 GeV and 1 TeV.

\begin{table*}[ht!]
  \centering
  \scalebox{0.95}{\begin{tabular}{|c|c|c|c|c|c|c|}
      \hline
      \multirow{ 3}{*}{Cuts}&\multicolumn{6}{c|}{Signals}\\\cline{2-7}
      &\multicolumn{2}{c}{$ee+\slashed{E}_T$}  &\multicolumn{2}{|c|}{$e\mu + \slashed{E}_T$} &\multicolumn{2}{c|}{$\mu\mu+\slashed{E}_T$}\\\cline{2-7}
       & $\sqrt{s}=0.5$ TeV &$\sqrt{s}=1.0$ TeV &$\sqrt{s}=0.5$ TeV &$\sqrt{s}=1.0$ TeV  & $\sqrt{s}=0.5$ TeV &$\sqrt{s}=1.0$ TeV\\\hline  
      Reject $E_{l_1} >$ [GeV]      &200 &350 &200 &300 &200 &350\\\hline
      Reject $E_{l_2} >$ [GeV]      &150 &300 &150 &300 &150 &300\\\hline
      Reject $M_{l_1l_2} >$ [GeV] &200 &200 &200 &250 &200 &200\\\hline
      Reject $\slashed{E}_T <$ [GeV] &270 &600 &270 &600 &270 &550\\\hline
  \end{tabular}}
  \caption{\label{table:a3}Proposed cuts for signal OSD$+\slashed{E}_T$.}
\end{table*}

\begin{table}[ht!]
  \centering
  \scalebox{0.95}{\begin{tabular}{|c|c|c|c|}
      \hline
      Cuts & BP1 (fb) &BP2 (fb) &Background (fb)\\
      \hline
      Preliminary &22.03 &2.46 &187.5\\
      \hline
      Reject $E_e >300$ GeV &21 &2.46 &51.47\\
      \hline
      Reject $E_\mu >300$ GeV &20.45 &2.46 &41.19\\
      \hline
      Reject $M_{e\mu} >250$ GeV &20.45 &2.46 &33.31\\
      \hline
      Reject $\slashed{E}_T <600$ GeV &18.55 &2.45 &16.56\\
      \hline 
  \end{tabular}}
  \caption{\label{table:a4}Cut flow of cross-section for BPs and background regarding signal $e^\pm \mu^\mp + \slashed{E}_T$ at $\sqrt{s}=1$ TeV.}
\end{table}

\section{Conclusion}
        {Dark matter and LFV searches are central to several investigations in the framework of the Scotogenic model. While opting for freeze-out scenarios for DM RD production, Yukawa couplings are relatively low to evade strong LFV bounds. We have investigated for the highest possible Yukawa coupling while satisfying the bounds from neutrino mass, dark matter relic density, LFV simultaneously. In this work, a freeze-out scenario is considered to generate the required DM RD  for the lightest $Z_2$--odd fermion being DM. We have explored collider signatures of our framework at future electron-positron colliders. Mono photons with missing energy and opposite sign dilepton with missing energy turn out to be two potential signatures to test our framework at the electron-positron colliders.}
        
\acknowledgments
We thank Saiyad Ashanujjaman for discussions. K.G. acknowledges the support from the SERB Core Research Grant [CRG/2019/006831].

\bibliographystyle{JHEP}
\bibliography{v0.bbl}


\end{document}